\documentclass[twocolumn,preprintnumbers,amsmath,amssymb,superscriptaddress,bibnotes, prb]{revtex4}
\usepackage{bm}% bold math
\usepackage{amssymb}
\usepackage{amsmath}
\usepackage{braket}
\usepackage{color}
\usepackage[separate-uncertainty = true,multi-part-units=single]{siunitx}
\usepackage{graphicx}% Include figure files
\newcommand{\beginsupplement}{%
        \setcounter{table}{0}
        \renewcommand{\thetable}{S\arabic{table}}%
        \setcounter{figure}{0}
        \renewcommand{\figurename}{Extended Data Fig.}
     }

\begin{document}
\title{Light-field driven currents in graphene}% Force line breaks with \\

\author{Takuya Higuchi} 
\affiliation{Chair of Laser Physics, Department of Physics, Friedrich-Alexander-Universit\"at Erlangen-N\"urnberg, Staudtstr. 1, D-91058 Erlangen, Germany}
\author{Christian Heide} 
\affiliation{Chair of Laser Physics, Department of Physics, Friedrich-Alexander-Universit\"at Erlangen-N\"urnberg, Staudtstr. 1, D-91058 Erlangen, Germany}
\author{Konrad Ullmann}
\affiliation{Chair of Applied Physics, Department of Physics, Friedrich-Alexander-Universit\"at Erlangen-N\"urnberg, Staudtstr. 7, D-91058 Erlangen, Germany}
\author{Heiko B. Weber}
\affiliation{Chair of Applied Physics, Department of Physics, Friedrich-Alexander-Universit\"at Erlangen-N\"urnberg, Staudtstr. 7, D-91058 Erlangen, Germany}
\author{Peter Hommelhoff}
\affiliation{Chair of Laser Physics, Department of Physics, Friedrich-Alexander-Universit\"at Erlangen-N\"urnberg, Staudtstr. 1, D-91058 Erlangen, Germany}

\maketitle

{\bf Ultrafast electron dynamics in solids under strong optical fields has recently found particular attention\cite{Ghimire2011,Schubert2014,Vampa2015,Langer2016,Schultze2012,Schultze2014,Schiffrin2012,Krausz2014,Golde2008}. In dielectrics and semiconductors, various light-field-driven effects have been explored, such as high-harmonic generation\cite{Ghimire2011,Schubert2014,Vampa2015,Langer2016}, sub-optical-cycle interband population transfer\cite{Schultze2012,Schultze2014} and nonperturbative increase of transient polarizability\cite{Schiffrin2012}. In contrast, much less is known about field-driven electron dynamics in metals because charge carriers screen an external electric field in ordinary metals\cite{Jackson,Schiffrin2012,Krausz2014}. Here we show that atomically thin monolayer Graphene offers unique opportunities to study light-field-driven processes in a metal. With a comparably modest field strength of up to 0.3 V/\AA, we drive combined interband and intraband electron dynamics, leading to a light-field-waveform controlled residual conduction current after the laser pulse is gone. We identify the underlying pivotal physical mechanism as electron quantum-path interference taking place on the 1-femtosecond (${\bf 10^{-15}}$ second) timescale. The process can be categorized as Landau-Zener-St\"uckelberg interferometry\cite{Shevchenko2010}. These fully coherent electron dynamics in graphene take place on a hitherto unexplored timescale faster than electron-electron scattering (tens of femtoseconds) and electron-phonon scattering (hundreds of femtoseconds)\cite{Breusing2011,Tielrooij2015,Johannsen2013,Gierz2013}. These results broaden the scope of light-field control of electrons in solids to an entirely new and eminently important material class -- metals -- promising wide ramifications for band structure tomography\cite{Schultze2014,Vampa2015} and light-field-driven electronics\cite{Krausz2014}.}

Graphene is an ideal platform to extend the concept of light-field-driven current control to metals. Even though the metallic nature of graphene is reflected in its excellent carrier mobilities\cite{CastroNeto2009,Novoselov2012}, the carrier concentration is low compared with conventional metals and thus screening due to free carriers is negligible at optical frequencies\cite{GarcadeAbajo2014}. Therefore, strong optical fields can be generated in graphene. In addition, graphene, in particular epitaxial graphene on SiC (0001), is one of the most robust materials available\cite{Emtsev2009,Novoselov2012}, and can thus withstand high laser intensities. Moreover, the optical response of graphene is broadband and ultrafast\cite{Novoselov2012}. Earlier photocurrent studies in graphene revealed that photocarriers are generated on an ultrashort timescale of tens of femtoseconds\cite{Breusing2011,Tielrooij2015}, associated with efficient and fast carrier heating\cite{Johannsen2013,Gierz2013,Malic2011}. Still, the timescale of these experiments is limited by the duration of the laser pulse (envelope) because the photocarrier generation is driven by optical absorption, which is governed by the cycle-averaged light intensity.  

\begin{figure*}
\begin{center}
\includegraphics[width=17cm]{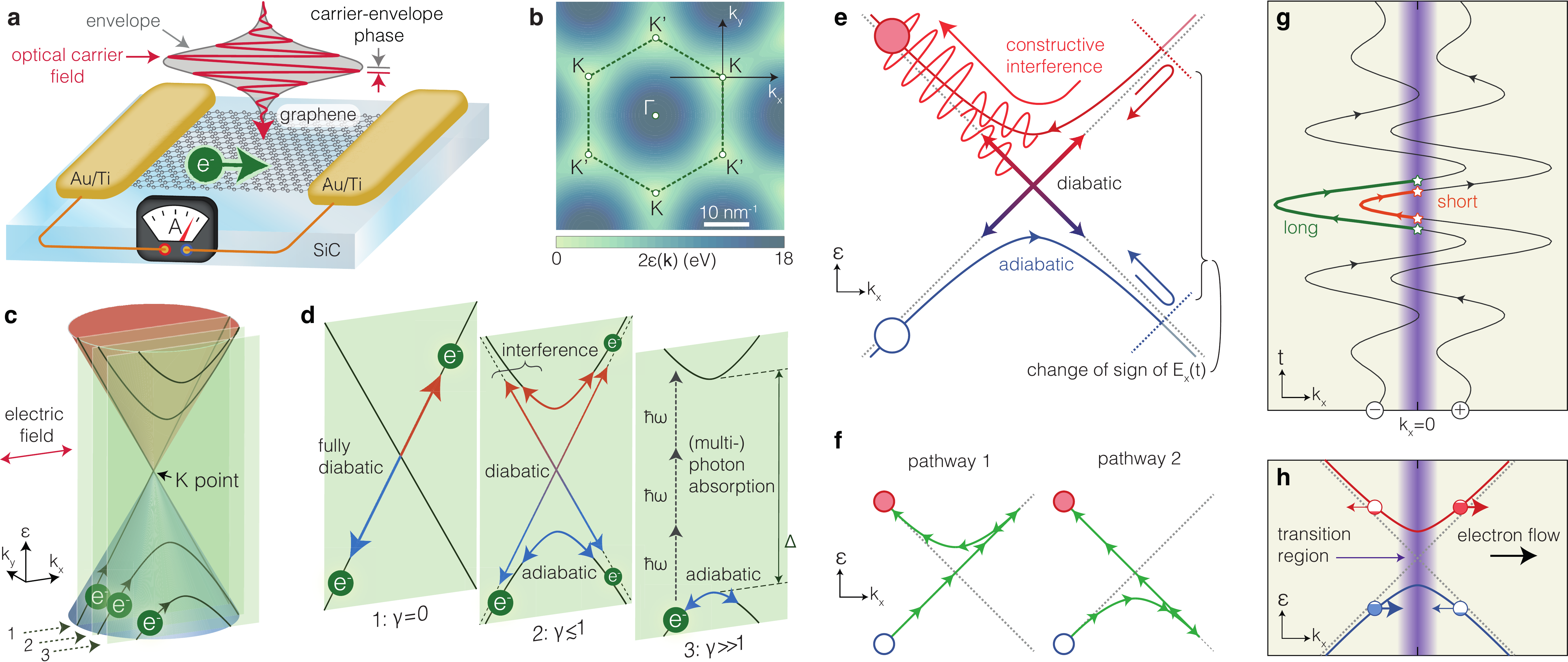}% Here is how to import EPS art
\caption{\label{Fig1} {\bf Overview of the experiment.} 
{\bf a}, A graphene stripe on a SiC substrate is illuminated with few-cycle carrier-envelope-phase stabilized laser pulses. The laser-induced current flowing through the graphene stripe, contacted with Au/Ti electrodes, is measured. 
{\bf b}, The electron dispersion relation of graphene. {\bf c}.
Around the $K$ (and $K'$) point, the dispersion shows the well-known gapless Dirac cone structure. The slope of the cone corresponds to the Fermi velocity $v_{\rm F}$. When electrons are driven by $x$-polarized light, only $k_x$ of the electron changes via acceleration, while $k_y$ does not. Hence the relevant dispersion is a slice of the cone with a fixed $k_y$. Three examples of such slices are shown in {\bf d}. A light-field driven electron undergoes a change of $k_x$.
How fast $v_{\rm F} k_x$ changes as a function of time compared with the apparent bandgap $\Delta$ determines the Keldysh adiabaticity parameter $\gamma$, leading to different excitation regimes. When $\gamma \ll 1$, the main contribution is diabatic interband transition, implying that the electron transition between valence and conduction bands. In particular, when $\gamma=0$ (slice 1), the transition is fully diabatic. When $\gamma \gg 1$ (slice 3), the adiabatic passage within one band is dominant and the diabatic transition is negligible. When $\gamma \approx 1$ (slice 2), the electron can undergo both diabatic transition and adiabatic passage. This combination of interband and intraband processes works as a beam splitter for the electron wave function. {\bf e}, In a single oscillation of the electric field, the electron experiences this beam splitting action twice, leading to two possible quantum pathways to reach the conduction band, {\bf f}, The electron starting from the valence band may either first pass through the beam splitter diabatically, and after the sign of the driving field has switched, adiabatically reach the conduction band (pathway 1), or {\it vice versa} (pathway 2). {\bf g}, Electron trajectories under the influence of a drive pulse with $\phi_{\rm CEP}=-\pi/2$, where the largest peak of the electron trajectory heads to negative $x$-direction. The change of the electron wave number is proportional to the vector potential of the driving field (see Methods for details). The purple area indicates the region where the interband transition probability is large. Stars represent the main interband transition events. {\bf h} Resulting asymmetric excitation probability that leads to the generation of the residual (persistent) current after the laser pulse is gone.
}
\end{center}
\end{figure*}

Here we show that a current induced in graphene by few-cycle laser pulses is sensitive to the electric-field waveform, i.e., the exact shape of the optical carrier {\it field} of the pulse, which is controlled by the carrier-envelope phase (Fig.~1a). As will be shown, the main mechanism of this waveform-dependent current generation is based on a large modulation of the interband coupling due to the intraband electron motion in reciprocal space, leading to quantum-path interference within a single optical cycle. Furthermore, Graphene's Dirac-cone dispersion relation (Figs.~1b, c) around the $K$ and $K'$ points (Dirac points)\cite{CastroNeto2009} is ideally suited to enable this interplay between the {\it interband} and {\it intraband} dynamics because of the steep change of the energy gap and of the dipole moment -- the two parameters determining the interband transition -- as a function of the wave number\cite{Ishikawa2013,Kelardeh2014}.

First, we consider electrons in graphene driven by an electric field ${\bf E}$ of linearly $x$-polarized light. Following the acceleration theorem ($\dot{\bf k} \propto {\bf E}$), only the $k_x$ component of the wave vector ${\bf k}$ is affected\cite{Grecchi2001}. Therefore, the relevant electron dispersion is a hyperbola formed by the intersection of a plane given by a fixed $k_y$ with the Dirac cone (Figs.~\ref{Fig1}c and d). This hyperbolic dispersion resembles an avoided crossing, and the size of the gap is determined by the $k_y$ value.

In planes far away from the Dirac point, a large apparent bandgap exists (slice ``3'' in Fig.~1d). Here, the intraband motion hardly affects interband transitions from the valence band to the conduction band. On the contrary, at the Dirac point, valence band and conduction band merge (slice ``1'' in Fig.~1d). Here, even small fields can drive electrons over the band crossing, and a full interband transition occurs instantaneously\cite{Ishikawa2013,Kelardeh2014}. Note that the electron returns to its original state after one optical cycle because it experiences this full transition twice: back and forth. Graphene, by virtue of its cone-shape energy dispersion, provides a continuous evolution between these two limiting cases. We note that the full span of the Keldysh model\cite{Keldysh1965} is covered between these two limits (see Methods): the Keldysh adiabaticity parameter $\gamma$ takes values $\gamma \gg 1$ for planes far away from the Dirac point (large $|k_y|$), and $\gamma \ll 1$ for those nearby (small $|k_y|$). 

An intermediate regime exists where both intraband motion and interband transitions appear in a coupled manner, as covered by the Landau-Zener framework\cite{Shevchenko2010}. In this intermediate regime, the avoided crossing may act as a beam splitter for electrons when they are driven by the external field and pass nearby the crossing: part of the electron wave function is transferred to the conduction band via interband diabatic transition, while the rest stays adiabatically in the valence band (intraband motion). The adiabatic motion leads to an adiabatic exchange of pseudo spins (the quantum number labeling the wave functions\cite{CastroNeto2009}), while in the diabatic process the pseudo spins are conserved.  
The intermediate regime corresponds to $\gamma \approx 1$ (slice ``2'' in Fig.~1d). While a single light pulse can drive all these cases with various $\gamma$ at once, the experimental scheme presented here is particularly sensitive to the intermediate regime. 

Since the electric field of light is of oscillating nature, electrons repeatedly pass nearby the avoided crossing even for a few-cycle laser pulse. This may result in different quantum pathways (Figs.~1e and f). These quantum pathways interfere, resulting in a conduction band population that depends on the phase relation between the pathways. This phase in turn is determined by the waveform of the applied laser field. Interference based on multiple Landau-Zener beam-splitting events is known as Landau-Zener-St\"uckelberg (LZS) interferometry\cite{Shevchenko2010}. LZS interferometry has so far only been observed in specially engineered quantum systems such as superconducting qubits\cite{Oliver2005}, cold atoms\cite{Kling2010}, or quantum dots\cite{Dovzhenko2011}. 

Even though the avoided crossing model is symmetric along the $k_x$ axis, the quantum path interference may result in an asymmetric population distribution after excitation with a few-cycle laser pulse. In particular, if the field-driven electron trajectory peaks toward the negative $x$-direction, the excursion between the two beam-splitting events is shorter for an initial $k_x>0$, compared with an initial $k_x<0$, leading to a difference in the relative phase between the two pathways (Fig.~1g). Therefore, the resultant conduction band population after the pulsed excitation may show an asymmetric distribution for $k_x>0$ and $k_x<0$ (Fig.~1h), leading to a residual current after the pulsed excitation\cite{Wismer2016}. This residual current is our observable.

The laser waveform, and hence the electron trajectory, is controlled by tuning the carrier-envelope phase (CEP, $\phi_{\rm CEP}$) of the few-cycle laser pulses (see Methods). Note that the spectral intensity and the intensity envelope are unchanged by variation of $\phi_{\rm CEP}$. We measure the CEP-dependent current through an unbiased graphene stripe under illumination of CEP-stabilized few-cycle laser pulses (Fig.~1a). The CEP-dependent light-induced current is extracted using a two-phase lock-in detection with the carrier-envelope offset frequency as the reference frequency. The difference in current between excitation pulses with $\phi_{\rm CEP}=0$ and $\pi$ is recorded as one of the lock-in signals $J_{\cos}$, and the one between $\phi_{\rm CEP} = \pi/2$ and $-\pi/2$ as $J_{\sin}$ (see Methods for details). We note that the laser spectrum is not octave-spanning, and so perturbative two-color interference effects\cite{Mele2000} do not contribute. 

\begin{figure}
\begin{center}
\includegraphics[width=8.5cm]{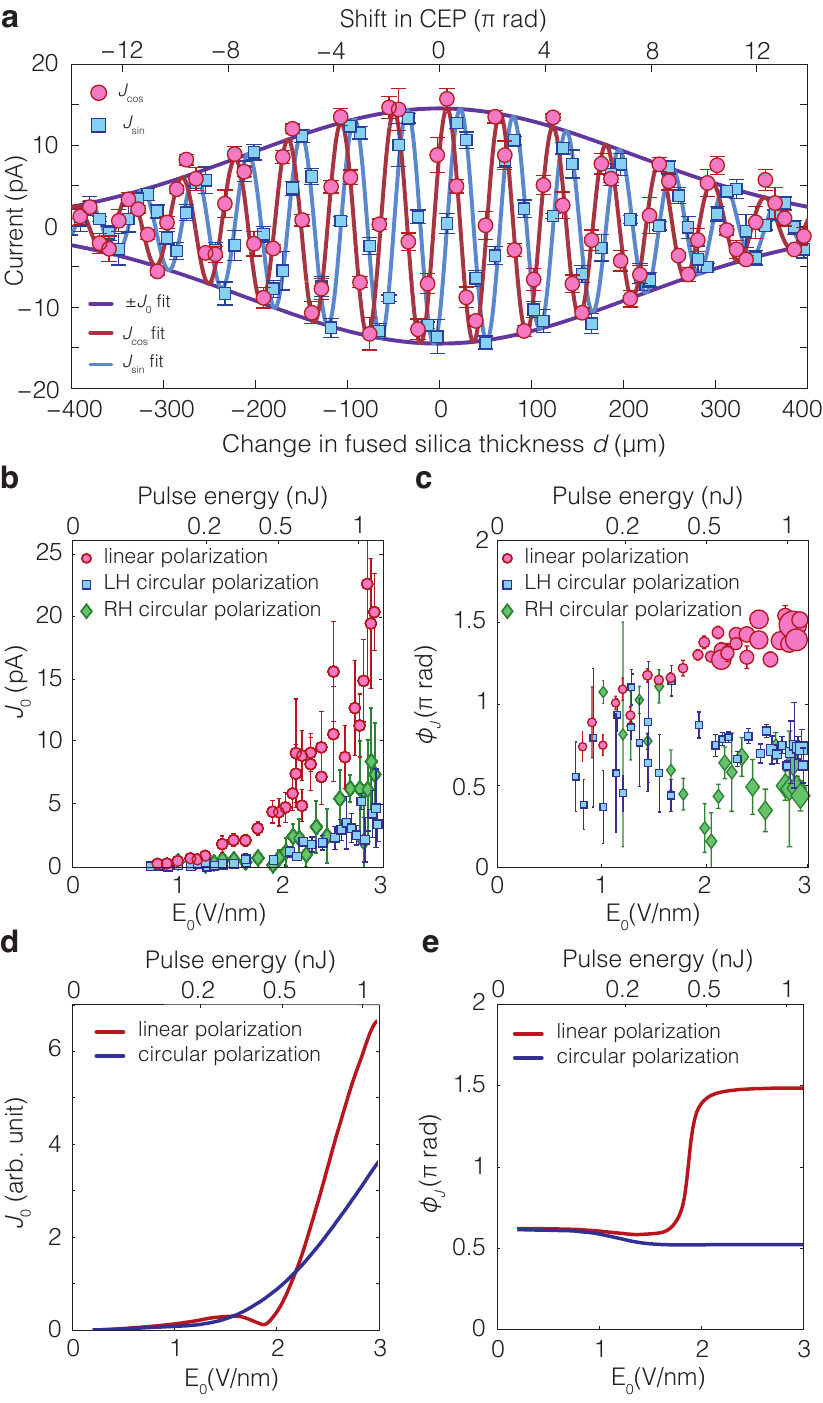}% Here is how to import EPS art
\caption{\label{Fig2} {\bf Measurement of the CEP-dependent current.} {\bf a} Carrier-envelope-phase-dependent current measured as two-phase lock-in signals $J_{\cos}$ (red markers, in phase) and $J_{\sin}$ (blue markers, quadrature) with the modulation reference frequency of $f_{\rm CEO}$, plotted as a function of relative thickness $d$ of fused silica in the beam path. The purple solid curve is a Gaussian fit of the signal amplitude $J_0$ as a function of $d$. Red and blue solid curves are fitting curves assuming this Gaussian fit of $J_0$ and the linear shift of the CEP at the sample position as a function of $d$ due to the spectral dispersion in fused silica. The optimal $d$ that maximizes $J_0$ is set to zero. {\bf b}, Amplitude $J_0$ and, {\bf c}, phase $\phi_{J}$ of the CEP-dependent lock-in current as a function of the peak laser field strength. (The size of the markers in {\bf c} also represents the amplitude $J_0$.) Results for linearly polarized parallel to the stripe direction, left-handed (LH) and right-handed (RH) circularly polarized excitations are plotted. Linearly polarized excitation {\it perpendicular} to the stripe direction does not generate a measurable CEP-dependent current at all ($<0.1$ pA, data not shown). 
{\bf d}, $J_0$ and, {\bf e}, $\phi_{J}$ obtained from the numerical simulations, as function of $E_0$. Simulated residual current after circularly polarized excitation does not depend on the handedness of the excitation.
Most importantly, the experimentally obtained phase change around $E_0 \approx$ 1.5 V/nm for linear polarization is also observed, meaning that the current changes its sign. The absence of such a phase change for circular polarization is also consistently observed in both experiment and simulation. 
} 
\end{center}
\end{figure}

Figure 2a shows the observed CEP-dependent current in graphene as a function of the (relative) thickness $d$ of fused silica in the beam path. At $d=0$, the pulse is the shortest, and the CEP-dependent current amplitude $J_0 \equiv\sqrt{J_{\cos}^2+J_{\sin}^2}$ is maximized under this condition.  As expected the phase $\phi_{J}$ of the lock-in signal (defined via $J_{\cos} = J_0 \cos\phi_{J}$ and $J_{\sin} = J_0 \sin\phi_{J}$) changes linearly as a function of $d$ and leads to oscillatory behaviours of $J_{\cos}$ and $J_{\sin}$ (see Methods for details), and clearly demonstrates the light-field-induced nature of the current. 

The CEP-dependent current amplitude $J_0$ at $d=0$ steeply increases as a function of the peak field strength $E_{0}$, as shown in Fig.~2b. The lock-in phase $\phi_{J}$ at $d=0$ increases as a function of $E_{0}$ for a light polarization parallel to the stripe direction (Fig.~2c). 
In total, $\phi_{J}$ is shifted by $0.8 \pi$ between $E_{0} < 1~{\rm V/nm}$ and $E_{0} > 2.8~{\rm V/nm}$. This means that the {\it direction} of the CEP-dependent current is almost reversed in between these two field strength regimes. On the other hand, circularly polarized excitation does not cause such a change in current direction. Pulses linearly polarized {\it perpendicular} to the graphene stripe do not generate a measurable CEP-dependent current at all ($<0.1$ pA). 

\begin{figure*}
\begin{center}
\includegraphics[width=17 cm]{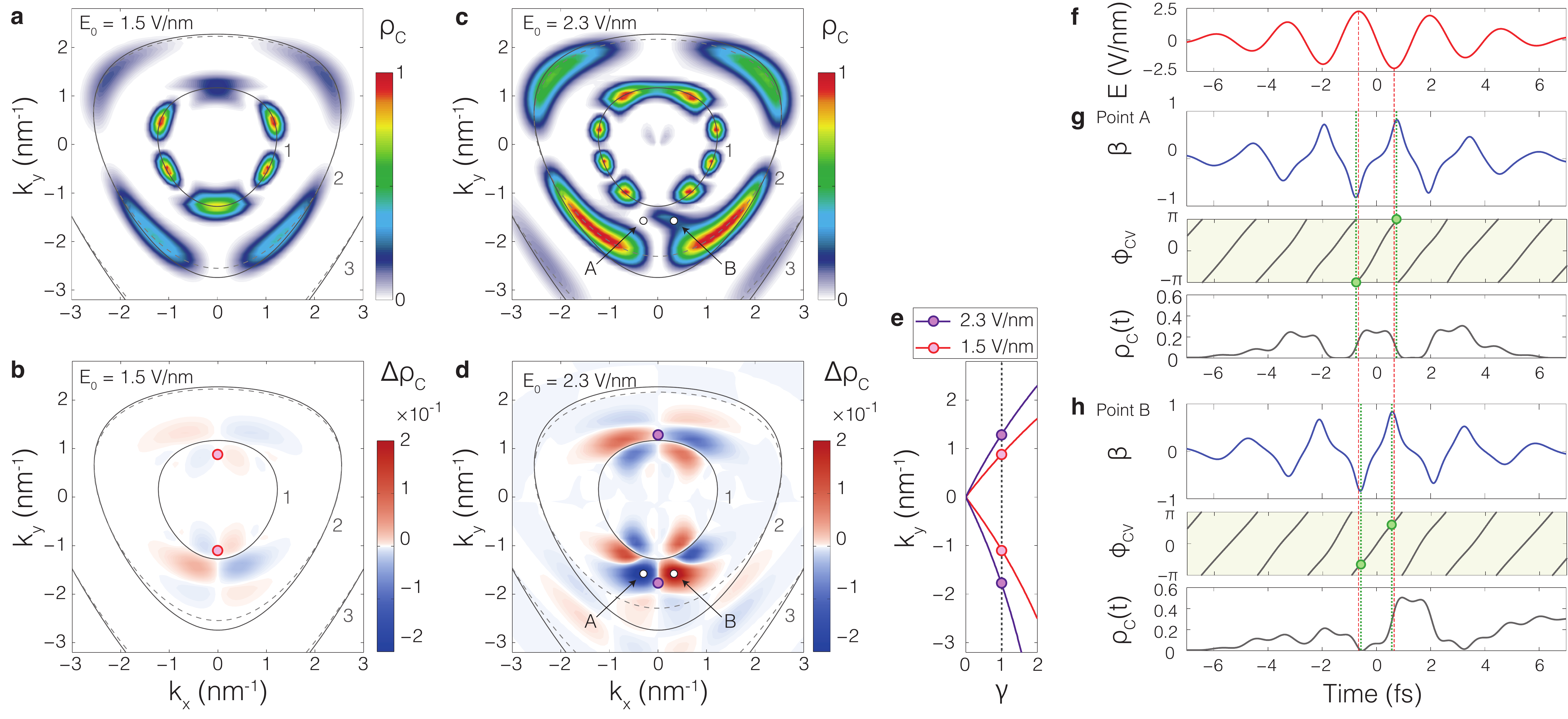}% Here is how to import EPS art
\caption{\label{Fig3} {\bf Quantum-path interference in CEP-dependent current generation with linearly polarized excitation.} {\bf a}, Simulated distribution of conduction-band population ${\rho}_{\rm C}$ after the laser excitation around the $K$ point with $E_0=1.5 ~{\rm V/nm}$ and $\phi_{\rm CEP}=\pi/2$. 
{\bf b}, Difference $\Delta{\rho}_{\rm C}$ between the respective ${\rho}_{\rm C}$ for excitations with $\phi_{\rm CEP}=\pi/2$ and $\phi_{\rm CEP}=-\pi/2$. {\bf c} and {\bf d} show the same as {\bf a} and {\bf b} but with $E_0=2.3~{\rm V/nm}$. 
Solid gray curves indicate wave vector values for which the energy difference between the two bands equals the mean laser photon energy, corresponding to resonant (multi-) photon absorption. The numbers of resonant photons required are indicated by the numbers.
{\bf e} Keldysh adiabaticity parameter $\gamma$ as a function of $k_y$ along $k_x =0$. Intersections with $\gamma=1$ are highlighted by markers. These pink and purple markers are duplicated in panels {\bf b} and {\bf d}. 
{\bf f} Electric field waveform of a pulse with $\phi_{\rm CEP} = \pi/2$ and $E_0=2.3~{\rm V/nm}$.
Relative band coupling strength $\beta(t)$, propagation phase $\phi_{\rm CV}(t)$ and conduction band population ${\rho}_{\rm C}(t)$ for the trajectory starting from point A ({\bf g}) and B ({\bf h}) in {\bf d}. Vertical red dashed lines indicate the driving fields' extrema. Vertical dash-dotted green lines indicate the maxima of $|\beta(t)|$. The green markers indicate the quantum phase $\phi_{\rm CV}$ for which $|\beta(t)|$ peaks. Note that the duration between these two green marker positions are longer than one half of an optical cycle  in {\bf g} but shorter in {\bf h}. This apparently small difference results in a qualitatively different interference condition. Notably, ${\rho}_{\rm C}(t)$ drops to zero at the second peak due to destructive interference in {\bf g}, while it increases due to constructive interference in {\bf h} (see text for details).
The repetitive passage of a Landau-Zener transition as observed and utilized here for the interpretation of the conduction band population and of the ensuing residual current is known as Landau-Zener-St\"uckelberg interferometry.
}
\end{center}
\end{figure*}

To clarify the origin of the CEP-dependent current and its peculiar dependence on the field strength, we model the dynamics of light-driven electrons in graphene within a nearest-neighbour tight-binding model plus laser-field interaction. The distribution of the conduction band population after the excitation by the laser pulse is numerically simulated, yielding the residual current (see Methods for details). The model reproduces the main features of the experiment well (Figs.~2d and e). The nonlinear increase of the CEP-dependent current as a function of field strength, as well as the striking change in current direction as a function of the peak field strength are fully reproduced. 

Figure 3a shows the simulated conduction-band population $\rho_{\rm C}$ after excitation with a linearly ($x$-) polarized pulse with $E_0 = 1.5 ~{\rm V/nm}$ and $\phi_{\rm CEP} = \pi/2$. Figure 3b shows the difference in $\rho_{\rm C}$ between excitation by a $\phi_{\rm CEP} = \pi/2$-pulse and a $\phi_{\rm CEP} = -\pi/2$-pulse. Figures 3c and 3d show the same for $E_0 = 2.3 ~{\rm V/nm}$. These results exhibit that several "hot spots" in $k$-space exist with clearly varying populations between the $\phi_{\rm CEP} = \pi/2$- and the $\phi_{\rm CEP} = -\pi/2$-excitations. These hot spots are distributed around $k_x \approx 0$, antisymmetric in the $k_x$-direction, and their $k_y$ values correspond to $\gamma \approx 1$ (Fig.~3e). For $E_0 = 1.5 ~{\rm V/nm}$, the main positive hot spot in Fig.~3b is located at $k_x<0$. This means that the $\phi_{\rm CEP} = \pi/2$-pulse creates more excitation for $k_x<0$, which corresponds to a flow of (negatively-charged) electrons in negative $x$-direction, i.e., a {\it positive} current in $x$-direction. On the other hand, for $E_0 = 2.3 ~{\rm V/nm}$, the positive main peak rests at $k_x>0$, leading to a current in {\it negative} $x$-direction, and directly evidencing the distinctive current change as function of field strength. As shown in Figs.~3b, d and e, the turning point coincides with the condition $\gamma = 1$ at the reciprocal points where the energy spacing between the two bands equals the one-photon energy (indicated by a distorted circle in Figs.~3a-d). A similar change in current direction has been theoretically predicted when the optical field amplitude cannot be treated as a perturbation, both in the case of the nonlinear photo-galvanic effect\cite{Wachter2015} and kicked anharmonic Rabi oscillations\cite{Wismer2016}, underscoring the non-perturbative nature of the light-matter interaction observed here.

A closer look on the temporal evolution of the interband transition probability provides insight into the role of the intra-optical-cycle LZS interference in the generation of the antisymmetric conduction band population distributions. 
The transition probability is governed by the ratio of the coupling between the two bands and the energy difference between them\cite{Shevchenko2010}. This ratio, which we call the relative band coupling strength, is given as $\beta({\bf k}_0, t) \equiv {\bf E}(t) \cdot {\bf d}({\bf k}(t)) [  \varepsilon({\bf k}(t))]^{-1}$, where ${\bf d}({\bf k})$ and $2\varepsilon({\bf k}(t))$ are the wavenumber-dependent dipole moment and the energy spacing, respectively. The LZS interference condition is governed by the two phases: the transition phase at a single Landau-Zener process (known as the Stokes phase) and the propagation phase between the two transition events\cite{Shevchenko2010}. The phase of $\beta$ determines the transition phase formed between initial and final state at the time of transition. In particular, a change in sign of $\beta$ means a difference of $\pi$ in the transition phase.
The propagation phase is the phase difference between the conduction- and valence-band states  and equals $\phi_{\rm CV}({\bf k}_0; t_1, t_2) \equiv \int _{t_1} ^{t_2} dt'  \hbar^{-1} 2\varepsilon({\bf k}(t'))$, where $t_1$ and $t_2$ refer to the transition events.

Figures~3g and h show $\beta({\bf k}_0, t)$, $\phi_{\rm CV}({\bf k}_0; 0, t)$ and the conduction band population ${\rho}_{\rm C}(t)$ as a function of time for two different initial ${\bf k}$-points labeled A and B in Fig.~3d, where the difference in $\rho_{\bf C}$ after excitation with a $\phi_{\rm CEP} = \pi/2$- vs. a $-\pi/2$-pulse is maximized. 
$\varepsilon({\bf k})$ and ${\bf d}({\bf k})$ are identical at points A and B. Therefore, if the intraband motion were negligible, the transition probability from the valence to the conduction bands should be the same. However, intraband motion leads to a difference in the LZS interference conditions for these two initial points.
In particular, the two main transition events (where $|\beta(t)|$ peaks, at $t_1$ and $t_2$) are more separated in time for the trajectory starting from point A, where $\phi_{\rm CV}({\bf k}_0;t_1,t_2) \approx 2\pi$. Together with the $\pi$-phase shift originating from the transition phase (i.e.~change in sign of $\beta(t)$), the total phase difference between the two LZS quantum pathways  is $\approx 3 \pi$, yielding destructive interference in the conduction band. On the other hand, for the electrons starting from point B, $\phi_{\rm CV}({\bf k}_0;t_1,t_2) \approx \pi$, which together with a transition phase difference of $\pi$ originating from the sign of $\beta$ results in constructive interference and thus a non-zero population in the conduction band. This difference in the quantum phase and hence the interference outcome leads to the asymmetric population distribution.

So far we have discussed one-dimensional trajectories of electrons along a line in reciprocal space, relevant for linear polarization: The electron travels back and forth in the same direction, allowing for two transition events within a single optical cycle. Thus quantum-path interference can play a dominant role. However, one can suppress this intra-optical-cycle interference by employing circular polarization. The absence of interference appears as the absence of the change of current direction with increasing field strength: $\phi_{J}$ is almost constant over the entire range of $E_0$ (Fig.~2c), well supported by simulation results (Fig.~2e). We interpret the CEP-dependent current for circular polarization as a result of the prominent contrast of the magnitude of the $|\beta|$-peaks between different CEPs. This is in qualitative difference to the case of linear polarization where the CEP does not influence the magnitude of the $|\beta|$ peaks but rather the spacings between them.

In summary, we have shown that in graphene a current can be generated that is sensitive to the carrier-envelope phase of few-cycle laser pulses. The main experimental features, especially the change in current direction as a function of field strength for linearly polarized excitation as well as the absence of such a change for circular polarization, are well reproduced by numerical simulations. These results can be interpreted with the presence and absence of sub-optical-cycle interference of electrons, known as Landau-Zener-St\"uckelberg interference. In the model, the electrons are treated fully coherently and independently because the sub-optical-cycle dynamics occurs on the 1-femtosecond timescale, and is thus faster than any scattering process\cite{Breusing2011,Johannsen2013,Gierz2013,Tielrooij2015}. If, in contrast, the dynamics are driven with more slowly oscillating fields such as those of terahertz pulses, the electron dynamics is rather incoherent due to collisions\cite{Tani2012}. Therefore, on an intermediate timescale, a door is now open to explore correlated electron dynamics on its fundamental timescale, as also indicated by quasi-particle collision studies in solids\cite{Langer2016}. The onset of these complex electron correlations might already cause the deviations between the experimental data and our initial model simulation results. In addition, the presence of other electronic bands is also predicted to influence the CEP-dependent photocurrent\cite{Wismer2016}, which will be important for future band structure tomography. We foresee that this demonstration of light-field driven currents in a low-dimensional metal paves the way to integrating electronics and optics on one platform.

\bibliography{../../Literature.bib}

\begin{thebibliography}{10}
\expandafter\ifx\csname url\endcsname\relax
  \def\url#1{\texttt{#1}}\fi
\expandafter\ifx\csname urlprefix\endcsname\relax\def\urlprefix{URL }\fi
\providecommand{\bibinfo}[2]{#2}
\providecommand{\eprint}[2][]{\url{#2}}

\bibitem{Ghimire2011}
\bibinfo{author}{Ghimire, S.} \emph{et~al.}
\newblock \bibinfo{title}{Observation of high-order harmonic generation in a
  bulk crystal}.
\newblock \emph{\bibinfo{journal}{Nat. Phys.}} \textbf{\bibinfo{volume}{7}},
  \bibinfo{pages}{138--141} (\bibinfo{year}{2011}).
\newblock \urlprefix\url{http://dx.doi.org/10.1038/nphys1847}.

\bibitem{Schubert2014}
\bibinfo{author}{Schubert, O.} \emph{et~al.}
\newblock \bibinfo{title}{Sub-cycle control of terahertz high-harmonic
  generation by dynamical {B}loch oscillations}.
\newblock \emph{\bibinfo{journal}{Nature Photonics}}
  \textbf{\bibinfo{volume}{8}}, \bibinfo{pages}{119--123}
  (\bibinfo{year}{2014}).
\newblock \urlprefix\url{http://dx.doi.org/10.1038/nphoton.2013.349}.

\bibitem{Vampa2015}
\bibinfo{author}{Vampa, G.} \emph{et~al.}
\newblock \bibinfo{title}{Linking high harmonics from gases and solids}.
\newblock \emph{\bibinfo{journal}{Nature}} \textbf{\bibinfo{volume}{522}},
  \bibinfo{pages}{462--464} (\bibinfo{year}{2015}).
\newblock \urlprefix\url{http://dx.doi.org/10.1038/nature14517}.

\bibitem{Langer2016}
\bibinfo{author}{Langer, F.} \emph{et~al.}
\newblock \bibinfo{title}{Lightwave-driven quasiparticle collisions on a
  subcycle timescale}.
\newblock \emph{\bibinfo{journal}{Nature}} \textbf{\bibinfo{volume}{533}},
  \bibinfo{pages}{225--229} (\bibinfo{year}{2016}).
\newblock \urlprefix\url{http://dx.doi.org/10.1038/nature17958}.

\bibitem{Schultze2012}
\bibinfo{author}{Schultze, M.} \emph{et~al.}
\newblock \bibinfo{title}{Controlling dielectrics with the electric field of
  light}.
\newblock \emph{\bibinfo{journal}{Nature}} \textbf{\bibinfo{volume}{493}},
  \bibinfo{pages}{75--78} (\bibinfo{year}{2012}).
\newblock \urlprefix\url{http://dx.doi.org/10.1038/nature11720}.

\bibitem{Schultze2014}
\bibinfo{author}{Schultze, M.} \emph{et~al.}
\newblock \bibinfo{title}{Attosecond band-gap dynamics in silicon}.
\newblock \emph{\bibinfo{journal}{Science}} \textbf{\bibinfo{volume}{346}},
  \bibinfo{pages}{1348--1352} (\bibinfo{year}{2014}).
\newblock \urlprefix\url{http://dx.doi.org/10.1126/science.1260311}.

\bibitem{Schiffrin2012}
\bibinfo{author}{Schiffrin, A.} \emph{et~al.}
\newblock \bibinfo{title}{Optical-field-induced current in dielectrics}.
\newblock \emph{\bibinfo{journal}{Nature}} \textbf{\bibinfo{volume}{493}},
  \bibinfo{pages}{70--74} (\bibinfo{year}{2012}).
\newblock \urlprefix\url{http://dx.doi.org/10.1038/nature11567}.

\bibitem{Krausz2014}
\bibinfo{author}{Krausz, F.} \& \bibinfo{author}{Stockman, M.~I.}
\newblock \bibinfo{title}{Attosecond metrology: from electron capture to future
  signal processing}.
\newblock \emph{\bibinfo{journal}{Nat. Photon.}} \textbf{\bibinfo{volume}{8}},
  \bibinfo{pages}{205--213} (\bibinfo{year}{2014}).
\newblock \urlprefix\url{http://dx.doi.org/10.1038/nphoton.2014.28}.

\bibitem{Golde2008}
\bibinfo{author}{Golde, D.}, \bibinfo{author}{Meier, T.} \&
  \bibinfo{author}{Koch, S.~W.}
\newblock \bibinfo{title}{High harmonics generated in semiconductor
  nanostructures by the coupled dynamics of optical inter- and intraband
  excitations}.
\newblock \emph{\bibinfo{journal}{Phys. Rev. B}} \textbf{\bibinfo{volume}{77}},
  \bibinfo{pages}{075330} (\bibinfo{year}{2008}).
\newblock \urlprefix\url{http://dx.doi.org/10.1103/PhysRevB.77.075330}.

\bibitem{Jackson}
\bibinfo{author}{Jackson, J.~D.}
\newblock \emph{\bibinfo{title}{Classical Electrodynamics Third Edition}}
  (\bibinfo{publisher}{Wiley}, \bibinfo{year}{1998}).

\bibitem{Shevchenko2010}
\bibinfo{author}{Shevchenko, S.}, \bibinfo{author}{Ashhab, S.} \&
  \bibinfo{author}{Nori, F.}
\newblock
  \bibinfo{title}{{L}andau{\textendash}{Z}ener{\textendash}{S}t\"{u}ckelberg
  interferometry}.
\newblock \emph{\bibinfo{journal}{Physics Reports}}
  \textbf{\bibinfo{volume}{492}}, \bibinfo{pages}{1--30}
  (\bibinfo{year}{2010}).
\newblock \urlprefix\url{http://dx.doi.org/10.1016/j.physrep.2010.03.002}.

\bibitem{Breusing2011}
\bibinfo{author}{Breusing, M.} \emph{et~al.}
\newblock \bibinfo{title}{Ultrafast nonequilibrium carrier dynamics in a single
  graphene layer}.
\newblock \emph{\bibinfo{journal}{Phys. Rev. B}} \textbf{\bibinfo{volume}{83}},
  \bibinfo{pages}{153410} (\bibinfo{year}{2011}).
\newblock \urlprefix\url{http://dx.doi.org/10.1103/PhysRevB.83.153410}.

\bibitem{Tielrooij2015}
\bibinfo{author}{Tielrooij, K.~J.} \emph{et~al.}
\newblock \bibinfo{title}{Generation of photovoltage in graphene on a
  femtosecond timescale through efficient carrier heating}.
\newblock \emph{\bibinfo{journal}{Nature Nanotech}}
  \textbf{\bibinfo{volume}{10}}, \bibinfo{pages}{437--443}
  (\bibinfo{year}{2015}).
\newblock \urlprefix\url{http://dx.doi.org/10.1038/NNANO.2015.54}.

\bibitem{Johannsen2013}
\bibinfo{author}{Johannsen, J.~C.} \emph{et~al.}
\newblock \bibinfo{title}{Direct view of hot carrier dynamics in graphene}.
\newblock \emph{\bibinfo{journal}{Phys. Rev. Lett.}}
  \textbf{\bibinfo{volume}{111}}, \bibinfo{pages}{027403}
  (\bibinfo{year}{2013}).
\newblock \urlprefix\url{http://dx.doi.org/10.1103/PhysRevLett.111.027403}.

\bibitem{Gierz2013}
\bibinfo{author}{Gierz, I.} \emph{et~al.}
\newblock \bibinfo{title}{Snapshots of non-equilibrium dirac carrier
  distributions in graphene}.
\newblock \emph{\bibinfo{journal}{Nature Materials}}
  \textbf{\bibinfo{volume}{12}}, \bibinfo{pages}{1119--1124}
  (\bibinfo{year}{2013}).
\newblock \urlprefix\url{http://dx.doi.org/10.1038/nmat3757}.

\bibitem{CastroNeto2009}
\bibinfo{author}{Neto, A. H.~C.}, \bibinfo{author}{Guinea, F.},
  \bibinfo{author}{Peres, N. M.~R.}, \bibinfo{author}{Novoselov, K.~S.} \&
  \bibinfo{author}{Geim, A.~K.}
\newblock \bibinfo{title}{The electronic properties of graphene}.
\newblock \emph{\bibinfo{journal}{Rev. Mod. Phys.}}
  \textbf{\bibinfo{volume}{81}}, \bibinfo{pages}{109--162}
  (\bibinfo{year}{2009}).
\newblock \urlprefix\url{http://dx.doi.org/10.1103/RevModPhys.81.109}.

\bibitem{Novoselov2012}
\bibinfo{author}{Novoselov, K.~S.} \emph{et~al.}
\newblock \bibinfo{title}{A roadmap for graphene}.
\newblock \emph{\bibinfo{journal}{Nature}} \textbf{\bibinfo{volume}{490}},
  \bibinfo{pages}{192--200} (\bibinfo{year}{2012}).
\newblock \urlprefix\url{http://dx.doi.org/10.1038/nature11458}.

\bibitem{GarcadeAbajo2014}
\bibinfo{author}{de~Abajo, F. J.~G.}
\newblock \bibinfo{title}{Graphene plasmonics: Challenges and opportunities}.
\newblock \emph{\bibinfo{journal}{{ACS} Photonics}}
  \textbf{\bibinfo{volume}{1}}, \bibinfo{pages}{135--152}
  (\bibinfo{year}{2014}).
\newblock \urlprefix\url{http://dx.doi.org/10.1021/ph400147y}.

\bibitem{Emtsev2009}
\bibinfo{author}{Emtsev, K.~V.} \emph{et~al.}
\newblock \bibinfo{title}{Towards wafer-size graphene layers by atmospheric
  pressure graphitization of silicon carbide}.
\newblock \emph{\bibinfo{journal}{Nature Materials}}
  \textbf{\bibinfo{volume}{8}}, \bibinfo{pages}{203--207}
  (\bibinfo{year}{2009}).
\newblock \urlprefix\url{http://dx.doi.org/10.1038/nmat2382}.

\bibitem{Malic2011}
\bibinfo{author}{Malic, E.}, \bibinfo{author}{Winzer, T.},
  \bibinfo{author}{Bobkin, E.} \& \bibinfo{author}{Knorr, A.}
\newblock \bibinfo{title}{Microscopic theory of absorption and ultrafast
  many-particle kinetics in graphene}.
\newblock \emph{\bibinfo{journal}{Phys. Rev. B}} \textbf{\bibinfo{volume}{84}},
  \bibinfo{pages}{205406} (\bibinfo{year}{2011}).
\newblock \urlprefix\url{http://dx.doi.org/10.1103/PhysRevB.84.205406}.

\bibitem{Ishikawa2013}
\bibinfo{author}{Ishikawa, K.~L.}
\newblock \bibinfo{title}{Electronic response of graphene to an ultrashort
  intense terahertz radiation pulse}.
\newblock \emph{\bibinfo{journal}{New J. Phys.}} \textbf{\bibinfo{volume}{15}},
  \bibinfo{pages}{055021} (\bibinfo{year}{2013}).
\newblock \urlprefix\url{http://dx.doi.org/10.1088/1367-2630/15/5/055021}.

\bibitem{Kelardeh2014}
\bibinfo{author}{Kelardeh, H.~K.}, \bibinfo{author}{Apalkov, V.} \&
  \bibinfo{author}{Stockman, M.~I.}
\newblock \bibinfo{title}{Wannier-stark states of graphene in strong electric
  field}.
\newblock \emph{\bibinfo{journal}{Phys. Rev. B}} \textbf{\bibinfo{volume}{90}},
  \bibinfo{pages}{085313} (\bibinfo{year}{2014}).
\newblock \urlprefix\url{http://dx.doi.org/10.1103/PhysRevB.90.085313}.

\bibitem{Grecchi2001}
\bibinfo{author}{Grecchi, V.} \& \bibinfo{author}{Sacchetti, A.}
\newblock \bibinfo{title}{Acceleration theorem for {B}loch oscillators}.
\newblock \emph{\bibinfo{journal}{Phys. Rev. B}} \textbf{\bibinfo{volume}{63}},
  \bibinfo{pages}{212303} (\bibinfo{year}{2001}).
\newblock \urlprefix\url{http://dx.doi.org/10.1103/PhysRevB.63.212303}.

\bibitem{Keldysh1965}
\bibinfo{author}{Keldysh, L.~V.}
\newblock \bibinfo{title}{Ionization in the field of a strong electromagnetic
  wave}.
\newblock \emph{\bibinfo{journal}{Sov. Phys. JETP}}
  \textbf{\bibinfo{volume}{20}}, \bibinfo{pages}{1307--1314}
  (\bibinfo{year}{1965}).

\bibitem{Oliver2005}
\bibinfo{author}{Oliver, W.~D.}
\newblock \bibinfo{title}{{M}ach-{Z}ehnder interferometry in a strongly driven
  superconducting qubit}.
\newblock \emph{\bibinfo{journal}{Science}} \textbf{\bibinfo{volume}{310}},
  \bibinfo{pages}{1653--1657} (\bibinfo{year}{2005}).
\newblock \urlprefix\url{http://dx.doi.org/10.1126/science.1119678}.

\bibitem{Kling2010}
\bibinfo{author}{Kling, S.}, \bibinfo{author}{Salger, T.},
  \bibinfo{author}{Grossert, C.} \& \bibinfo{author}{Weitz, M.}
\newblock \bibinfo{title}{Atomic {B}loch-{Z}ener oscillations and
  {S}t\"{u}ckelberg interferometry in optical lattices}.
\newblock \emph{\bibinfo{journal}{Phys. Rev. Lett.}}
  \textbf{\bibinfo{volume}{105}}, \bibinfo{pages}{215301}
  (\bibinfo{year}{2010}).
\newblock \urlprefix\url{http://dx.doi.org/10.1103/PhysRevLett.105.215301}.

\bibitem{Dovzhenko2011}
\bibinfo{author}{Dovzhenko, Y.} \emph{et~al.}
\newblock \bibinfo{title}{Nonadiabatic quantum control of a semiconductor
  charge qubit}.
\newblock \emph{\bibinfo{journal}{Phys. Rev. B}} \textbf{\bibinfo{volume}{84}},
  \bibinfo{pages}{161302} (\bibinfo{year}{2011}).
\newblock \urlprefix\url{http://dx.doi.org/10.1103/PhysRevB.84.161302}.

\bibitem{Wismer2016}
\bibinfo{author}{Wismer, M.~S.}, \bibinfo{author}{Kruchinin, S.~Y.},
  \bibinfo{author}{Ciappina, M.}, \bibinfo{author}{Stockman, M.~I.} \&
  \bibinfo{author}{Yakovlev, V.~S.}
\newblock \bibinfo{title}{Strong-field resonant dynamics in semiconductors}.
\newblock \emph{\bibinfo{journal}{Phys. Rev. Lett.}}
  \textbf{\bibinfo{volume}{116}}, \bibinfo{pages}{197401}
  (\bibinfo{year}{2016}).
\newblock \urlprefix\url{http://dx.doi.org/10.1103/PhysRevLett.116.197401}.

\bibitem{Mele2000}
\bibinfo{author}{Mele, E.~J.}, \bibinfo{author}{Kr{\'{a}}l, P.} \&
  \bibinfo{author}{Tom{\'{a}}nek, D.}
\newblock \bibinfo{title}{Coherent control of photocurrents in graphene and
  carbon nanotubes}.
\newblock \emph{\bibinfo{journal}{Phys. Rev. B}} \textbf{\bibinfo{volume}{61}},
  \bibinfo{pages}{7669--7677} (\bibinfo{year}{2000}).
\newblock \urlprefix\url{http://dx.doi.org/10.1103/PhysRevB.61.7669}.

\bibitem{Wachter2015}
\bibinfo{author}{Wachter, G.} \emph{et~al.}
\newblock \bibinfo{title}{Controlling ultrafast currents by the nonlinear
  photogalvanic effect}.
\newblock \emph{\bibinfo{journal}{New Journal of Physics}}
  \textbf{\bibinfo{volume}{17}}, \bibinfo{pages}{123026}
  (\bibinfo{year}{2015}).
\newblock \urlprefix\url{http://stacks.iop.org/1367-2630/17/i=12/a=123026}.

\bibitem{Tani2012}
\bibinfo{author}{Tani, S.}, \bibinfo{author}{Blanchard, F.} \&
  \bibinfo{author}{Tanaka, K.}
\newblock \bibinfo{title}{Ultrafast carrier dynamics in graphene under a high
  electric field}.
\newblock \emph{\bibinfo{journal}{Phys. Rev. Lett.}}
  \textbf{\bibinfo{volume}{109}}, \bibinfo{pages}{166603}
  (\bibinfo{year}{2012}).
\newblock \urlprefix\url{http://dx.doi.org/10.1103/PhysRevLett.109.166603}.

\bibitem{Landau1932}
\bibinfo{author}{Landau, L.}
\newblock \bibinfo{title}{Zur theorie der energieubertragung. ii}.
\newblock \emph{\bibinfo{journal}{Phys. Z. Sowjetunion}}
  \textbf{\bibinfo{volume}{2}}, \bibinfo{pages}{46--51} (\bibinfo{year}{1932}).

\bibitem{Zener1932}
\bibinfo{author}{Zener, C.}
\newblock \bibinfo{title}{Non-adiabatic crossing of energy levels}.
\newblock \emph{\bibinfo{journal}{Proc. R. Soc. A}}
  \textbf{\bibinfo{volume}{137}}, \bibinfo{pages}{696--702}
  (\bibinfo{year}{1932}).
\newblock \urlprefix\url{http://dx.doi.org/10.1098/rspa.1932.0165}.

\bibitem{Ishikawa2010}
\bibinfo{author}{Ishikawa, K.~L.}
\newblock \bibinfo{title}{Nonlinear optical response of graphene in time
  domain}.
\newblock \emph{\bibinfo{journal}{Phys. Rev. B}} \textbf{\bibinfo{volume}{82}},
  \bibinfo{pages}{201402} (\bibinfo{year}{2010}).
\newblock \urlprefix\url{http://dx.doi.org/10.1103/PhysRevB.82.201402}.

\bibitem{Holthaus2000}
\bibinfo{author}{Holthaus, M.}
\newblock \bibinfo{title}{{B}loch oscillations and {Z}ener breakdown in an
  optical lattice}.
\newblock \emph{\bibinfo{journal}{J Opt B Quantum Semiclassical Opt}}
  \textbf{\bibinfo{volume}{2}}, \bibinfo{pages}{589--604}
  (\bibinfo{year}{2000}).
\newblock \urlprefix\url{http://dx.doi.org/10.1088/1464-4266/2/5/306}.

\bibitem{Tielrooij2013}
\bibinfo{author}{Tielrooij, K.~J.} \emph{et~al.}
\newblock \bibinfo{title}{Photoexcitation cascade and multiple hot-carrier
  generation in graphene}.
\newblock \emph{\bibinfo{journal}{Nat. Phys.}} \textbf{\bibinfo{volume}{9}},
  \bibinfo{pages}{248--252} (\bibinfo{year}{2013}).
\newblock \urlprefix\url{http://dx.doi.org/10.1038/nphys2564}.

\bibitem{Chen2015}
\bibinfo{author}{Chen, J.-J.} \emph{et~al.}
\newblock \bibinfo{title}{Photovoltaic effect and evidence of carrier
  multiplication in graphene vertical homojunctions with asymmetrical metal
  contacts}.
\newblock \emph{\bibinfo{journal}{{ACS} Nano}} \textbf{\bibinfo{volume}{9}},
  \bibinfo{pages}{8851--8858} (\bibinfo{year}{2015}).
\newblock \urlprefix\url{http://dx.doi.org/10.1021/acsnano.5b02356}.

\end{thebibliography}

%% Here is the endmatter stuff: Supplementary Info, etc.
%% Use \item's to separate, default label is "Acknowledgements"

\noindent {\bf [Acknowledgement]} This work has been supported in part by the European Research Council (ERC NearFieldAtto) and Deutsche Forschungsgemeinschaft (SFB 953).

\noindent {\bf [Author Contributions]} T.H. and P.H. conceived the study. T.H. and C.H. conveyed the current measurement experiments, analyzed the data, and provided the plots. K.U. fabricated the sample under supervision of H.B.W. T.H. conveyed the numerical simulation. T.H. and P.H. wrote the manuscript with the input from all the authors. All authors discussed the obtained results.

\noindent {\bf [Competing Interests]} The authors declare that they have no competing financial interests.

\noindent {\bf [Correspondence]} Correspondence and requests for materials should be addressed to  \\ T.H.~(email: takuya.higuchi@fau.de) or \\
P.H.  (email: peter.hommelhoff@fau.de).

%%
%% TABLES
%%
%% If there are any tables, put them here.
%%
\beginsupplement
\newpage

\noindent{\bf Sample preparation.}
Monolayer graphene is grown epitaxially on a step-bunched 6H-SiC(0001) substrate\cite{Emtsev2009}. A graphene stripe with a width of $\SI{2.0 \pm 0.1}{\micro \metre}$ is patterned on a single terrace step by electron beam lithography and plasma etching. Two gold electrodes (thickness of 50 nm) with titanium adhesive layers (thickness of 5 nm) are deposited on the stripe, with a distance of $\SI{4.0 \pm 0.1}{\micro \metre}$ between them (Extended Data Figs.~1a, b and c).

\begin{figure}
\begin{center}
\includegraphics[width=8.5 cm]{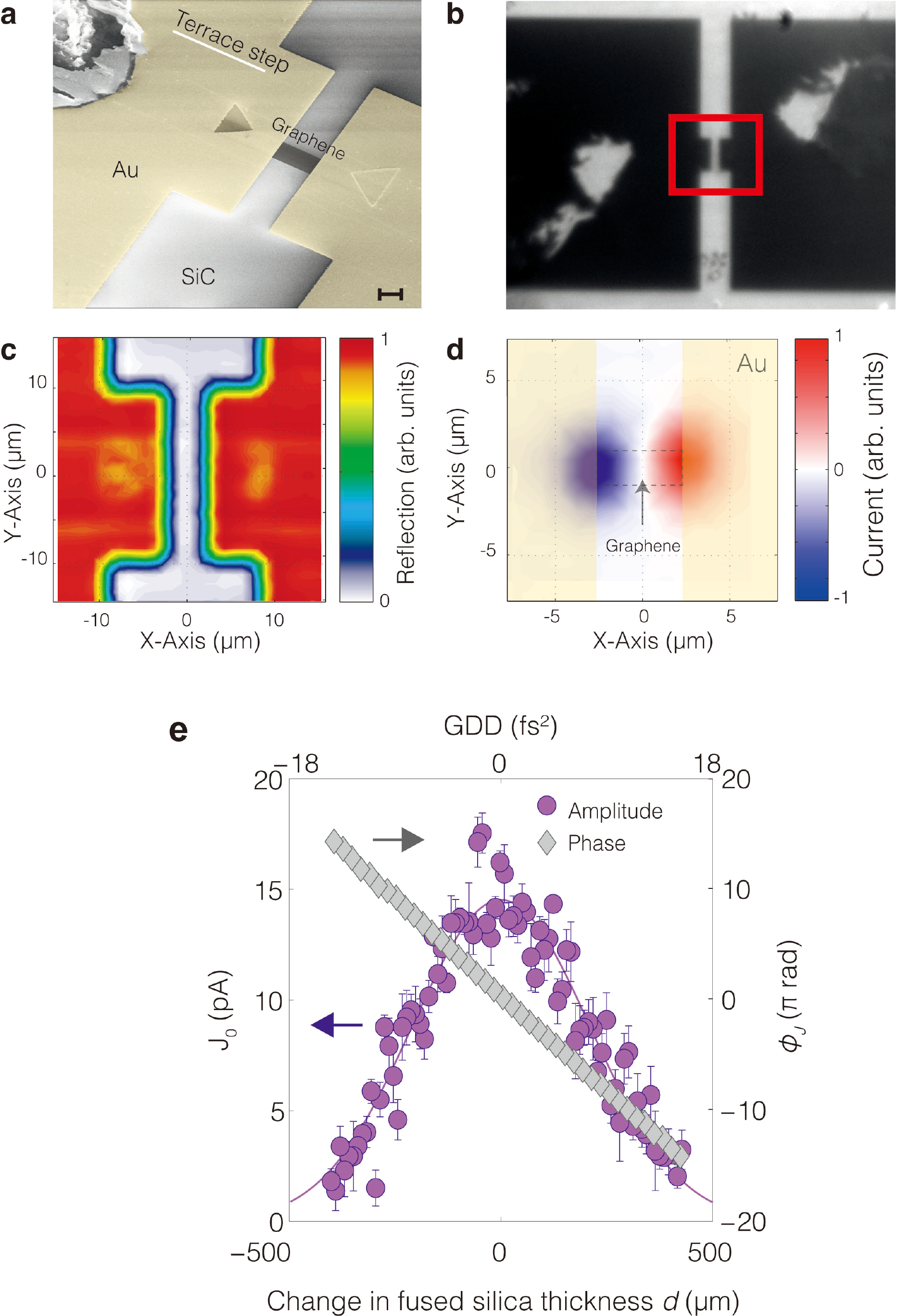}% Here is how to import EPS art
\caption{{\bf Sample geometry and laser alignment} {\bf a.} Scanning-electron-microscope image of the sample. Scale bar is $\SI{2}{\micro \metre}$. {\bf b.} Transmission optical microscope image of the sample. The dark areas correspond to the metal electrodes. {\bf c.} Map of reflected light intensity while scanning the laser spot two-dimensionally. The scanning area is indicated by the red box in {\bf b}.
The metal electrodes provide larger reflection. {\bf d.} Map of (CEP-independent) photocurrent while scanning the laser spot. When the laser spot hits the graphene-metal junction, photo carriers are generated in graphene and they result in photocurrent due to the build-in potential at the junction originating from the mismatch of the work functions. {\bf e.} Amplitude $J_0$ and phase $\phi_J$ of the CEP-dependent current as a function of fused silica insertion thickness $d$. The data is same as Fig.~2a, but the phase is shown instead of the in-phase component $J_{\cos}$ and the quadrature component $J_{\sin}$.}
\end{center}
\end{figure}

The graphene is $n$-doped, with a carrier concentration of $n=(8.0\pm0.9)\times10^{12}$ cm$^{-1}$ and a carrier mobility of $\mu=(860\pm 60)$ cm$^{2}\cdot$V$^{-1}\cdot $s$^{-1}$, determined by Hall and conductivity measurements. The corresponding Fermi energy is $E_{\rm F}=0.3$ eV, which implies that the plasmonic response of electrons (and thus screening) is negligible at the photon energies ($\approx 1.5$ eV) of the laser employed\cite{GarcadeAbajo2014}.

\vspace{10 pt}
\noindent{\bf Laser system.}
A laser oscillator (VENTEON) with a repetition rate of 80 MHz, a central wavelength of 800 nm, a Fourier-limited pulse duration of 5.4 fs (i.e., FWHM of the intensity envelope) is used as the source of laser. The laser pulses are focused to the middle of the graphene stripe by an off-axis parabolic mirror, and the spot size is $\SI{1.6\pm0.1}{\micro \metre}$ ($1/e^2$ intensity radius). This spot size is much smaller than the distance between the two electrodes.

At the junctions between graphene and the electrodes, built-in fields are formed due to the mismatch of the work functions. This built-in fields produce a CEP-independent photocurrent when the junctions are illuminated. We minimize this CEP-independent photocurrent by aligning the laser position to the middle of the graphene stripe (Extended Data Fig.~1d).

Since the graphene is placed on the surface of a dielectric (SiC), the optical field strength applied to graphene is reduced compared with that in the bare focus in vacuum by a factor $\frac{2}{1+n_{\rm SiC}}$, where $n_{\rm SiC} \approx 2.6$ is the refractive index of SiC at the laser center wavelength. The field strength $E_0$ includes this reduction.

\vspace{10 pt}
\noindent{\bf Measurement of the CEP-dependent current.}
The carrier-envelope offset frequency of the laser is stabilized with a home-built $f$-$2f$ interferometer. Long-term drifts in the CEP are corrected using an out-of-loop $f$-$2f$ interferometer.

The optically induced current through the stripe is measured with a current amplifier (Stanford Research Systems 570). We extract the CEP-dependent current by a two-phase lock-in amplifier (Stanford Research Systems 810), with the carrier-envelope offset frequency $f_{\rm CEO}$ as the reference frequency. 

When the $f_{\rm CEO}$ is nonzero, the CEP $\phi_{\rm CEP}(t)$ of the laser pulses slips in time (from pulse to pulse), $\phi_{\rm CEP}(t) = 2\pi f_{\rm CEO} t + \phi_0$. Here the additional phase offset $\phi_0$ is determined by the group delay dispersion of the dispersive medium in the beam path. 
The two-phase lock-in amplifier detects the in-phase component $J_{\cos}$ and the quadrature component $J_{\sin}$. The amplitude $J_0\equiv\sqrt{J_{\cos}^2+J_{\sin}^2}$ is the magnitude of the CEP-dependent current. The angle $\phi_{J}$  is defined by $J_{\cos}=J_0 \cos(\phi_{J})$ and $J_{\sin}=J_0 \sin(\phi_{J})$.

By inserting a transparent but dispersive material (such as fused silica) with thickness $d$ to the beam path, two things happen to the pulse. The first is the overall shift of the CEP at the sample position by $\phi_0 = d (v_{\rm p}^{-1}-v_{\rm g}^{-1})$, where $v_{\rm p}$ and $v_{\rm g}$ are phase and group velocities of light in the material at the laser central wavelength, respectively. This shift in the CEP at the sample position reveals itself in change of $\phi_{J}$, as shown in Extended Data Fig.~1e. The second is the change in the pulse duration via the dispersion, and thus the peak field strength is maximized for an optimal glass thickness $d$. Hence, the lock-in signal amplitude $J_0$ shows a peak as a function of $d$.

We have calibrated $\phi_{J}$ by shifting the $\phi_{J}$ axis of the experimental data so that the $\phi_{J}$ values at $d=0$ and $E_0>2.5$ V/nm coincide with the numerically simulated results. The same shift is used for all the experimental conditions (with variations of field strengths and polarizations), and thus the phase relation between the experimental data is maintained.

\vspace{10 pt}
\noindent{\bf Waveform and polarization.}
The electric field of linearly polarized light is described as $E_x(t) = E_0 \alpha(t) \cos(\omega t + \phi_{\rm CEP})$, where $E_0$ is the peak field amplitude, $\omega$ is the central frequency, $\phi_{\rm CEP}$ is the carrier-envelope phase, and $\alpha(t) $ is the normalized envelope function with a maximum $\alpha=1$ at $t=0$.

To generate circularly polarized light, a broadband quarter wave plate is placed into the beam path. By rotating the optical axis of this waveplate by $\pm 45$ degrees regarding the polarization direction of the incident linearly polarized light, the laser pulses become circularly polarized. The electric field vector of a circularly polarized laser pulse is 
\begin{eqnarray}
E_x(t) &=& \frac{1}{\sqrt{2}} E_0 \alpha(t) \cos(\omega t + \phi_{\rm CEP}), \\
E_y(t) &=& \pm \frac{1}{\sqrt{2}} E_0 \alpha(t) \sin(\omega t + \phi_{\rm CEP}), 
\end{eqnarray}
where the sign $\pm$ represents if the helicity is left or right.

\vspace{10 pt}
\noindent{\bf Adiabaticity parameter.}
As discussed in the main text, by considering an intersecting plane of a Dirac cone with a constant $k_y$, the dispersion relation has the form of an avoided crossing. 
This avoided crossing model has been employed to describe if the intraband motion may affect an interband transition by means of the adiabaticity parameter $\gamma$ introduced by Keldysh\cite{Keldysh1965}. 
The original form of the adiabaticity parameter $\gamma$ for an avoided-crossing model is\cite{Keldysh1965}
\begin{equation}
\gamma \equiv \frac{\omega \sqrt{m  \Delta}}{|e|E_{0}} \label{eq-Keldysh},
\end{equation}
where $\omega$ and $E_{0}$ are the angular frequency and the amplitude of the oscillating electric field, $m$ is the reduced mass of the electron and the hole, $m^{-1} = m_{\rm e}^{-1} + m_{\rm h}^{-1}$, and $ \Delta$ is the band gap. 

Now we apply this formula to the Dirac Hamiltonian of graphene. The Hamiltonian is ${\cal H}({\bf k}) = \hbar v_{\rm F} \sigma \cdot {\bf k}$, where $\sigma$ is the two-by-two Pauli matrix and $v_{\rm F}$ is the slope of the Dirac-cone dispersion. 
Consider a case of $x$-polarized excitation, where an intersecting plane with a fixed ${k_y}$ is relevant as discussed in the main text. Within this plane, the eigenvalues of this Hamiltonian are
\begin{equation}
\pm \varepsilon({\bf k}) =\pm  \hbar v_{\rm F} \sqrt{ k_x^2 + k_y^2 } = \pm \hbar v_{\rm F} |k_y| \sqrt{1 + \frac{k_x^2}{k_y^2}}.
\end{equation}
From this dispersion relation, one can obtain the apparent band gap
$ \Delta$ and the effective mass $m$ as
\begin{eqnarray}
 \Delta &=& 2\varepsilon (k_x=0, k_y) = 2 \hbar v_{\rm F} k_y,\\
 m &=& \hbar^2 \left( \frac{\partial^2 (2\varepsilon ({\bf k})) }{\partial k_x^2} \right)_{k_x=0}^{-1} = \frac{\hbar |k_y|}{2v_{\rm F}}.
\end{eqnarray}
By substituting these results to Eq.~\eqref{eq-Keldysh}, we obtain
\begin{equation}
\gamma = \frac{\hbar |k_y| \omega }{|e|E_{0}}.
\end{equation}

The Keldysh adiabaticity parameter $\gamma$ can be interpreted as the ratio between the minimum band gap $2\varepsilon(k_x=0, k_y) = 2\hbar v_{\rm F} |k_y|$ and the maximum energy difference $2\hbar v_{\rm F}|e|E_{0}\omega^{-1}$ of electrons in the two original (crossing) bands originating from the intraband motion. Therefore, if the electron's initial wavenumber is nearby the $K$ point (where $2\varepsilon(k_x=0, k_y) \approx 0$), 
$\gamma \lesssim 1$ can be satisfied easily and the change in band gap due to the intraband motion is not negligible compared with $2\varepsilon(k_x=0, k_y)$. In this case, the large change in wave number induces diabatic transition of electrons between the valence and the conduction bands akin to Landau-Zener tunneling\cite{Landau1932,Zener1932}. Note that when $\gamma \ll 1$, the transition probability converges to the Landau-Zener formula\cite{Keldysh1965}. In particular, when the electron trajectory passes through the $K$ point, then $\gamma = 0$, and the electron dynamics is purely diabatic\cite{Ishikawa2010,Kelardeh2014}. On the other hand, when $\gamma \gg 1$, electron dynamics is well described by the vertical inter-band transition, such as (multi-) photon absorption. Because of the conical dispersion relation of graphene, a wide range of $\gamma$ is spanned, and the CEP-dependent excitation is mainly found at $\gamma \approx 1$ in the numerical simulation.

\begin{figure}
\begin{center}
\includegraphics[width=8.5cm]{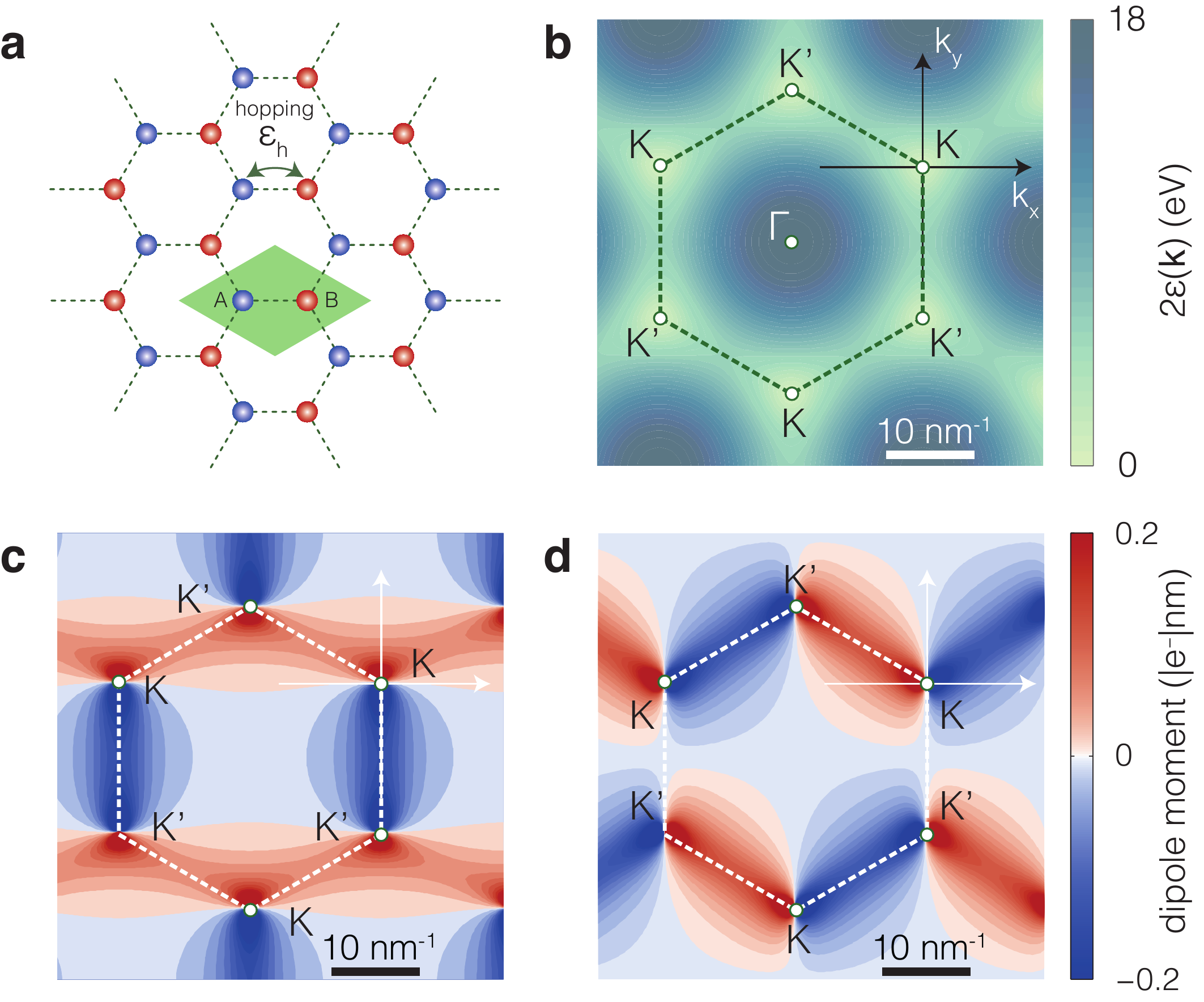}% Here is how to import EPS art
\caption{{\bf Theoretical model for simulation} {\bf a.} Lattice structure of graphene. There are two carbon sites A and B in a unit cell (green shaded area). {\bf b.} Energy difference between the conduction and the valence band obtained from the tight-binding model. {\bf c.} and {\bf d.} are $x$ and $y$ components of the dipole matrix element ${\bf d}({\bf k})$.}
\end{center}
\end{figure}

\vspace{10 pt}
\noindent{\bf Band structure of graphene.}
We consider a tight-binding model for graphene (Extended Data Fig.~2a) with a hopping parameter of $\varepsilon_{\rm h}=3.0$ eV between nearest neighboring atoms described by the Hamiltonian:
\begin{equation}
{\cal H}_0 =  -\varepsilon_{\rm h} \sum_{<i,j>}  a_i^\dagger b_j + b_j^\dagger a_i.
\end{equation}
The positions of the nearest-neighbor sites around a carbon atom at sublattice (a) are
\begin{equation}
{\bf b}_1 = \frac{a}{\sqrt{3}} {\bf e}_x, ~~
{\bf b}_{2,3} = \frac{a}{\sqrt{3}} \left( -\frac{1}{2} {\bf e}_x \pm \frac{\sqrt{3}}{2} {\bf e}_y \right),
\end{equation}
where $a = 0.246$ nm is the lattice constant of graphene.

For the basis functions,
\begin{eqnarray}
\phi_{\bf k}^{(a)} &=& \frac{1}{\sqrt{N}} \sum_{n=1}^N e^{i {\bf k}\cdot{\bf r}} \phi \left({\bf r}-{\bf r}^{\rm (a)}_{n} \right), \nonumber \\
\phi_{\bf k}^{(b)} &=& \frac{1}{\sqrt{N}} \sum_{n=1}^N e^{i {\bf k}\cdot{\bf r}} \phi \left( {\bf r}-{\bf r}^{\rm (b)}_{n} \right), \label{eq-AandBbasis}
\end{eqnarray}
the Hamiltonian is
\begin{equation}
{\cal H}_0 = \sum _{\bf k} - \varepsilon_{\rm h} \left( f({\bf k}) \alpha_{\bf k}^{(a)\dagger } \alpha_{\bf k}^{(b) } + f^*({\bf k}) \alpha_{\bf k}^{(b)\dagger } \alpha_{\bf k}^{(a) }\right),
\end{equation}
where
\begin{equation}
f({\bf k}) = \exp \left( i \frac{a k_x}{\sqrt{3}} \right) + 2 \exp \left( -i \frac{a k_x}{2\sqrt{3}} \right) \cos \left(\frac{a k_y}{2} \right).
\end{equation}
Here, $\alpha_{\bf k}^{(a)}$ and $\alpha_{\bf k}^{(b)}$ are the annihilation operators for an electron with the wave function $\phi_{\bf k}^{(a)} $  and $\phi_{\bf k}^{(b)} $, respectively. 
We assume that the interaction of an electron with the other electrons or phonons can be neglected due to the short timescale of the light-electron interaction considered here, and thus the interaction between electrons with different initial ${\bf k}$ values are neglected. Therefore, one can introduce a matrix form to describe a single ${\bf k}$ value of this Hamiltonian as:
\begin{equation}
{\cal H}_0({\bf k}) = \left[ \begin{array}{cc} 
0 & - \varepsilon_{\rm h}   f({\bf k}) \\ 
- \varepsilon_{\rm h}  f^*({\bf k})  & 0 \end{array} \right].
\end{equation}
Note that we neglect the overlap integral between different atomic sites, as well as the transfer to other than the nearest neighbor sites. One can introduce these effects into the theory, but they do not lead to qualitative changes.

The Hamiltonian is diagonalized with a unitary matrix $U_{\bf k}$ defined as
\begin{eqnarray}
U^\dagger_{\bf k} &\equiv& \frac{1} {\sqrt{2}} \left[ \begin{array}{cc} e^{-i \theta_{\bf k}/2} & e^{i \theta_{\bf k}/2} \\ -e^{-i \theta_{\bf k}/2} & e^{i \theta_{\bf k}/2} \end{array} \right], \nonumber \\
U_{\bf k} &\equiv& \frac{1} {\sqrt{2}} \left[ \begin{array}{cc} 
e^{i \theta_{\bf k}/2} & -e^{i \theta_{\bf k}/2} \\ 
e^{-i \theta_{\bf k}/2} & e^{-i \theta_{\bf k}/2} \end{array} \right],
\end{eqnarray}
\begin{equation}
U^\dagger_{\bf k}{\cal H}_{0}({\bf k}) U_{\bf k} = 
\left[
\begin{array}{cc}
-\varepsilon_{\rm h} |f({\bf k})| & 0 \\ 0 & \varepsilon_{\rm h} |f({\bf k})|
\end{array}
\right]
\end{equation}
where the angle $\theta_{\bf k}$ is defined as 
\begin{equation}
e^{i\theta_{\bf k}} \equiv \frac{f({\bf k})}{|f({\bf k})|}.
\end{equation}

The corresponding basis functions are the conduction band and valence band wave functions defined as
\begin{equation}
\phi_{\bf k}^{\rm (v)} ({\bf r}) = \frac{1} {\sqrt{2}} \left( \begin{array}{c} e^{i \theta_{\bf k}/2} \\ e^{-i \theta_{\bf k}/2} \end{array} \right) \label{eq-wavefunctionv}
\end{equation}
and
\begin{equation}
\phi_{\bf k}^{\rm (c)} ({\bf r}) = \frac{1}{\sqrt{2}} \left( \begin{array}{c} -e^{i \theta_{\bf k}/2} \\ e^{-i \theta_{\bf k}/2} \end{array} \right)
,
\end{equation}
respectively. The corresponding eigenenergies are $\pm \varepsilon_{\rm h} |f({\bf k})|$ (Extended Data Fig.~2b).
Next we calculate the dipole matrix element between the conduction and valence states:
\begin{equation}
{\bf d}({\bf k}) = \bra{\phi_{\bf k}^{\rm (v)}} e {\bf r} \ket{\phi_{\bf k}^{\rm (c)}} = \frac{e}{2} \nabla_{\bf k} \theta_{\bf k}. \label{eq-DipoleMoment}
\end{equation}
The $x$ and $y$ components of the dipole matrix element are
\begin{equation}
d_x({\bf k}) = \frac{ea}{2\sqrt{3}} \frac{1 + \cos\left( \frac{a k_y}{2} \right) \left[ \cos \left( \frac{\sqrt{3}ak_x}{2} \right) - 2 \cos \left( \frac{ak_y}{2} \right) \right] }{1 + 4 \cos\left( \frac{a k_y}{2} \right) \left[ \cos\left( \frac{\sqrt{3}ak_x}{2} \right) + \cos \left( \frac{ak_y}{2} \right) \right] },
\end{equation}
\begin{equation}
d_y({\bf k}) = \frac{ea}{2} \frac{ \sin\left(\frac{\sqrt{3}ak_x}{2}\right) \sin\left(\frac{ak_y}{2}\right) }{1 + 4 \cos\left( \frac{a k_y}{2} \right) \left[ \cos\left( \frac{\sqrt{3}ak_x}{2} \right) + \cos \left( \frac{ak_y}{2} \right) \right] },
\end{equation}
as shown in Extended Data Figs.~2c and d.

\vspace{10 pt}
\noindent{\bf Dynamics of electrons in graphene under an optical field.}
An external electric field drives electrons via the dipole interaction, introduced via ${\cal H}_{\rm I}$, and thus the total Hamiltonian ${\cal H}(t)$ reads 
\begin{equation}
{\cal H}(t) = {\cal H}_0 + {\cal H}_{\rm I} \equiv {\cal H}_0 - e {\bf E}(t) \cdot {\bf r} . \label{eq-BareHamiltonian}
\end{equation}

Based on the acceleration theorem \cite{Golde2008,Holthaus2000,Grecchi2001}, the wave number of electrons in the reciprocal space changes as $\dot{\bf k} = \hbar^{-1} e {\bf E}(t) $. In the integrated form, ${\bf k} = \int_{-\infty}^t \hbar^{-1} e {\bf E}(t') dt' = {\bf k}_0 - \hbar^{-1} e {\bf A}(t) $, where ${\bf k}_0$ is the initial wavenumber and ${\bf A}(t)$ is the vector potential. We choose the Coulomb gauge. Therefore, it is useful to construct an ansatz involving this acceleration to solve the time-dependent Schr\"odinger equation including the change in wavenumber of electrons. Hence, we start from the ansatz:
\begin{equation}
\phi_{{\bf k}_0}(t) = c^{(a)}_{{\bf k}_0}(t) \phi^{(a)}_{ {\bf k}(t)} + c^{(b)}_{{\bf k}_0}(t) \phi^{(b)}_{ {\bf k}(t)},
\end{equation} 
where $\phi^{(a)}_{{\bf k}(t)}$ and $\phi^{(b)}_{{\bf k}(t)}$ (cf. Eq.~\eqref{eq-AandBbasis}) are the basis functions, and now the wave number ${\bf k}(t)$ changes as a function of time.

Temporal evolution of this ansatz is as follows:
\begin{eqnarray}
& & i\hbar \frac{\partial}{\partial t} \phi_{{\bf k}_0}(t) \nonumber \\
&=& i \hbar \big(  \dot c^{(a)}_{{\bf k}_0}(t) \phi^{(a)}_{ {\bf k}(t)} + c^{(a)}_{{\bf k}_0}(t) \dot \phi^{(a)}_{ {\bf k}(t)} \nonumber \\ 
&& ~~~~~~~~~~~ + \dot c^{(b)}_{{\bf k}_0}(t) \phi^{(b)}_{ {\bf k}(t)} + c^{(b)}_{{\bf k}_0}(t) \dot \phi^{(b)}_{ {\bf k}(t)} \big) \nonumber \\ 
&=& i\hbar \left( \dot c^{(a)}_{{\bf k}_0}(t) \phi^{(a)}_{ {\bf k}(t)} + \dot c^{(b)}_{{\bf k}_0}(t) \phi^{(b)}_{ {\bf k}(t)} \right) + i \hbar {\frac{\partial {\bf k}(t)}{\partial t}} \cdot \nabla_{\bf k} \phi_{{\bf k}_0}(t) \nonumber \\
&=& i\hbar \left( \dot c^{(a)}_{{\bf k}_0}(t) \phi^{(a)}_{ {\bf k}(t)} + \dot c^{(b)}_{{\bf k}_0}(t) \phi^{(b)}_{ {\bf k}(t)} \right) -  e {\bf E}(t) \cdot {\bf r} \phi_{\bf k}(t), \label{eq-TemporalEvol}
\end{eqnarray}
and the last term is the same as the interaction Hamiltonian.
Here we use the relation $\dot \phi^{(a)}_{ {\bf k}(t)} = i \hbar {\frac{\partial {\bf k}(t)}{\partial t}} \cdot \left(\nabla_{\bf k} \phi^{(a)}_{ \bf k}\right)_{\bf k = {\bf k}(t)}$ and $\dot \phi^{(b)}_{ {\bf k}(t)} = i \hbar {\frac{\partial {\bf k}(t)}{\partial t}} \cdot \left(\nabla_{\bf k} \phi^{(b)}_{ \bf k}\right)_{\bf k = {\bf k}(t)}$ because the ${\bf k}$-dependence of $\phi^{(a)}_{\bf k}$ and $\phi^{(b)}_{\bf k}$ is only introduced at the plane wave part in Eq.~\eqref{eq-AandBbasis}. By comparing Eq.~\eqref{eq-BareHamiltonian} and Eq.~\eqref{eq-TemporalEvol}, we obtain
\begin{equation}
i\hbar \left( \dot c^{(a)}_{{\bf k}_0}(t) \phi^{(a)}_{ {\bf k}(t)} + \dot c^{(b)}_{{\bf k}_0}(t) \phi^{(b)}_{ {\bf k}(t)} \right) = {\cal H}_0({\bf k}={\bf k}(t)) \phi_{\bf k}(t).
\end{equation}

Then the temporal evolution of the coefficients $c^{(a)}_{{\bf k}_0}(t)$ and $c^{(b)}_{{\bf k}_0}(t)$ is:
\begin{eqnarray}
&&\frac{\partial}{\partial t} 
\left[ \begin{array}{c}
c^{(a)}_{{\bf k}_0}(t) \\ c^{(b)}_{{\bf k}_0}(t)
\end{array} \right] \nonumber \\
&=& 
\frac{1}{i \hbar }
\left[ \begin{array}{cc} 
0 & -\varepsilon_{\rm h}   f \left( {\bf k}(t) \right) \\ 
-\varepsilon_{\rm h}  f^* \left( {\bf k}(t) \right)  & 0 \end{array} \right]
\left[ \begin{array}{c}
c^{(a)}_{{\bf k}_0}(t) \\ c^{(b)}_{{\bf k}_0}(t)
\end{array} \right].\label{eq-OpticalInteraction}
\end{eqnarray} 
Let $\tilde {\cal H}_{\bf k}(t)$ be the matrix in the right-hand side.

We simulated the temporal evolution of the conduction band population distribution on the basis of Eq.~\eqref{eq-OpticalInteraction}. The equation of motion is discritized in time and then numerically integrated with the Crank-Nicolson method:
\begin{eqnarray}
&&\left[ \begin{array}{c}
c^{(a)}_{{\bf k}_0}(t_{n+1}) \\ c^{(b)}_{{\bf k}_0}(t_{n+1})
\end{array} \right] \nonumber \\
&=& \left( {\bf I} +  \frac{ \Delta_t \tilde{\cal H}_{\bf k}(t_{n+1})}{2i\hbar} \right)^{-1}
\left( {\bf I} -  \frac{\Delta_t \tilde{\cal H}_{\bf k}(t_{n})}{2i\hbar} \right)
\left[ \begin{array}{c}
c^{(a)}_{{\bf k}_0}(t_{n}) \\ c^{(b)}_{{\bf k}_0}(t_{n})
\end{array} \right]. \nonumber \\
\end{eqnarray}

After calculating the temporal evolution of the factors $c^{(a)}_{{\bf k}_0}(t)$ and $ c^{(b)}_{{\bf k}_0}(t)$, we convert these values to conduction and valence band populations by the following relations:
\begin{equation}
c^{(a)}_{{\bf k}_0}(t) \phi^{(a)}_{ {\bf k}(t)} + c^{(b)}_{{\bf k}_0}(t) \phi^{(b)}_{ {\bf k}(t)} = c^{(v)}_{{\bf k}_0}(t) \phi^{(v)}_{ {\bf k}(t)} + c^{(c)}_{{\bf k}_0}(t) \phi^{(c)}_{ {\bf k}(t)},
\end{equation}
or, in the matrix form:
\begin{equation}
\left[ \begin{array}{c}
c^{(v)}_{{\bf k}_0}(t) \\ c^{(c)}_{{\bf k}_0}(t)
\end{array} \right]
=
U^\dagger_{ {\bf k}(t) }
\left[ \begin{array}{c}
c^{(a)}_{{\bf k}_0}(t) \\ c^{(b)}_{{\bf k}_0}(t)
\end{array} \right].
\end{equation}
The conduction band population is $\rho^{(c)}_{{\bf k}_0}(t) = |c^{(c)}_{{\bf k}_0}(t)|^2$.

Note that it is also possible to obtain the temporal evolution of the linear combination coefficients $c^{(v)}_{{\bf k}_0}(t)$ and $c^{(c)}_{{\bf k}_0}(t)$, where the conduction and the valence bands are taken as the basis functions. The equation of motion for these coefficients is
\begin{eqnarray}
&&\frac{\partial}{\partial t}
\left[ \begin{array}{c}
c^{(v)}_{{\bf k}_0}(t) \\ c^{(c)}_{{\bf k}_0}(t)
\end{array} \right] \nonumber \\
&=&
\frac{\partial}{\partial t}
\left(
U^\dagger_{ {\bf k}(t) }
\left[ \begin{array}{c}
c^{(a)}_{{\bf k}_0}(t) \\ c^{(b)}_{{\bf k}_0}(t)
\end{array} \right]
\right)  \nonumber \\
&=& 
\frac{\partial}{\partial t}
\left(
U^\dagger_{ {\bf k}(t) }\right)
\left[ \begin{array}{c}
c^{(a)}_{{\bf k}_0}(t) \\ c^{(b)}_{{\bf k}_0}(t)
\end{array} \right]
+
U^\dagger_ {  {\bf k}(t) } 
\frac{\partial}{\partial t}
\left[ \begin{array}{c}
c^{(a)}_{{\bf k}_0}(t) \\ c^{(b)}_{{\bf k}_0}(t)
\end{array} \right]  \nonumber \\
&=& 
\frac{\partial}{\partial t}
\left(
U^{\dagger}_{ {\bf k}(t) }\right)
U_{ {\bf k}(t) }
\left[ \begin{array}{c}
c^{(v)}_{{\bf k}_0}(t) \\ c^{(c)}_{{\bf k}_0}(t)
\end{array} \right] \nonumber \\
&&~~~~~~~~~~~~~+
U^{\dagger}_ {  {\bf k}(t) } 
\frac{1}{i\hbar}
{\cal H}_0( {\bf k}(t))
\left[ \begin{array}{c}
c^{(a)}_{{\bf k}_0}(t) \\ c^{(b)}_{{\bf k}_0}(t)
\end{array} \right]  \nonumber \\
&=& 
\left(
\frac{\partial}{\partial t}
\left(
U^{\dagger}_{ {\bf k}(t) }\right)
U_{ {\bf k}(t) }
+
\frac{1}{i\hbar}
U^{\dagger}_ {  {\bf k}(t) } 
{\cal H}_0( {\bf k}(t))
U_ {  {\bf k}(t) } 
\right)
\left[ \begin{array}{c}
c^{(v)}_{{\bf k}_0}(t) \\ c^{(c)}_{{\bf k}_0}(t)
\end{array} \right]. \nonumber \\ 
\label{eq-EqMotion}
\end{eqnarray}
The first term in the last parenthesis is 
\begin{eqnarray}
&&\frac{\partial}{\partial t}
\left(
U^{\dagger}_{ {\bf k}(t) }\right)
U_{ {\bf k}(t) } \nonumber \\
&=& 
\frac{\partial {\bf k}(t) }{\partial t} \cdot \left( \nabla_{\bf k}  U^{\dagger}_{ \bf k } \right) _{ \bf k = {\bf k}(t)} U_{{\bf k}(t)} \nonumber \\
&=& 
\frac{ e{\bf E}(t) }{\hbar} \cdot 
\left(\frac{i}{2}\nabla_{\bf k}\theta_{\bf k} \right)_{\bf k = {\bf k}(t)} 
\left[
\begin{array}{cc}
0 & 1 \\ 1 & 0
\end{array}
\right]  \nonumber \\
&=& \frac{1}{i\hbar} \left[
\begin{array}{cc}
0 & -{\bf E}(t)\cdot {\bf d}({\bf k}={\bf k}(t)) \\ -{\bf E}(t)\cdot {\bf d}({\bf k}={\bf k}(t)) & 0
\end{array}
\right], \nonumber \\
\end{eqnarray}
which represents the coupling between the conduction and the valence band via the interaction of the electric field and the dipole between the two bands.

The second term in the last parenthesis of Eq.~\eqref{eq-EqMotion} is the diagonalized Hamiltonian at each ${\bf k}={\bf k}(t)$, and the diagonal values are the eigenenergies of the conduction and valence band electrons.

In total, the temporal evolution of the coefficients is described as
\begin{eqnarray}
&& \frac{\partial}{\partial t}
\left[ \begin{array}{c}
c^{(v)}_{{\bf k}_0}(t) \\ c^{(c)}_{{\bf k}_0}(t)
\end{array} \right] \nonumber \\
&=&
\frac{1}{i\hbar} \left[
\begin{array}{cc}
-\varepsilon_{\rm h} |f({\bf k}(t))| & -{\bf E}(t)\cdot {\bf d}({\bf k}(t)) \\ -{\bf E}(t)\cdot {\bf d}({\bf k}(t)) & \varepsilon_{\rm h} |f({\bf k}(t))|
\end{array}
\right] 
\left[ \begin{array}{c}
c^{(v)}_{{\bf k}_0}(t) \\ c^{(c)}_{{\bf k}_0}(t)
\end{array} \right]. \nonumber \\ 
\label{eq-OpticalBloch}
\end{eqnarray}
This gives the equation of motion of conduction-band and valence-band electrons in graphene coupled with light via electric dipole interaction.
The diagonal terms describe the energy of the valence and conduction bands, and the off-diagonal terms describe the dipole coupling. 

The two equations \eqref{eq-OpticalInteraction} and \eqref{eq-OpticalBloch} are equivalent. Therefore, one can integrate both equations to obtain the same temporal evolution. However, in the case of graphene, the conduction- and valence-band wavefunctions, and thus the dipole moment, have singularities at the $K$ and $K'$ points. Therefore, the numerical integration of the Eq.~\eqref{eq-OpticalBloch} requires careful treatment around these singular points\cite{Ishikawa2010,Ishikawa2013,Kelardeh2014}, and hence we take an approach based on Eq.~\eqref{eq-OpticalInteraction}.

\vspace{10 pt}
\noindent{\bf Estimation of the residual current after the interaction.}
To estimate the current from the residual conduction band population distribution $\rho_{\rm C}({\bf k})$, it is necessary to consider propagation effects afterwards. For this, we introduce two assumptions. The first is that the carriers propagate with the group velocities given as the slope of the dispersion at the electron wave numbers. The second is that the carriers are multiplied with a multiplication factor proportional to the initial carrier energy $\varepsilon ({\bf k})=\varepsilon_{\rm h} |f({\bf k}(t))|$ just after the excitation \cite{Tielrooij2013,Chen2015}. Due to the conical band structure around the $K$ ($K'$) points, the most probable multiplication is forward scattering that does not change the group velocity\cite{Tielrooij2013,Chen2015}. Therefore, the residual current $J$ satisfies:
\begin{equation}
J \propto \int_{{\rm BZ},\varepsilon ({\bf k})>E_{\rm F}} d{\bf k} \rho_{\rm C}({\bf k}) \frac{\partial \varepsilon ({\bf k})}{\partial {\bf k}} \varepsilon ({\bf k}),
\end{equation}
where the integral is taken over the Brillouin zone (BZ) of graphene with a condition $\varepsilon ({\bf k})>E_{\rm F}$ because conduction band states below the Fermi energy $E_{\rm F}$ are initially occupied before the excitation and cannot be accessed via interband transitions due to Pauli Blocking.

To compare the simulation results with the experimental data (Figs.~2b and 2c), the electric field waveform is calculated from the experimentally obtained laser spectrum, assuming a flat spectral phase.
We assume the beam spot shape of a two-dimensional Gaussian function with a $1/e^2$ radius of $\SI{1.6}{\micro \metre}$, and the current is integrated over the beam spot. 

The simulation results in Figs.~3 and 4 are obtained by assuming a Gaussian temporal envelope with a full width at half maximum of $5.4$ fs, which represents the main feature of the electric field waveform. The CEP-dependence in the excitation probability is highly nonlinear in the laser intensity, and thus the analysis of the main temporal peak is sufficient. The experimental intensity spectrum contains several peaks, which results in minor but complex structures in the conduction band population distribution when it is used as the input for the simulation. These minor structures qualitatively do not influence the total current after integrating over the BZ and the beam spot. 

\begin{figure}
\begin{center}
\includegraphics[width=8.9cm]{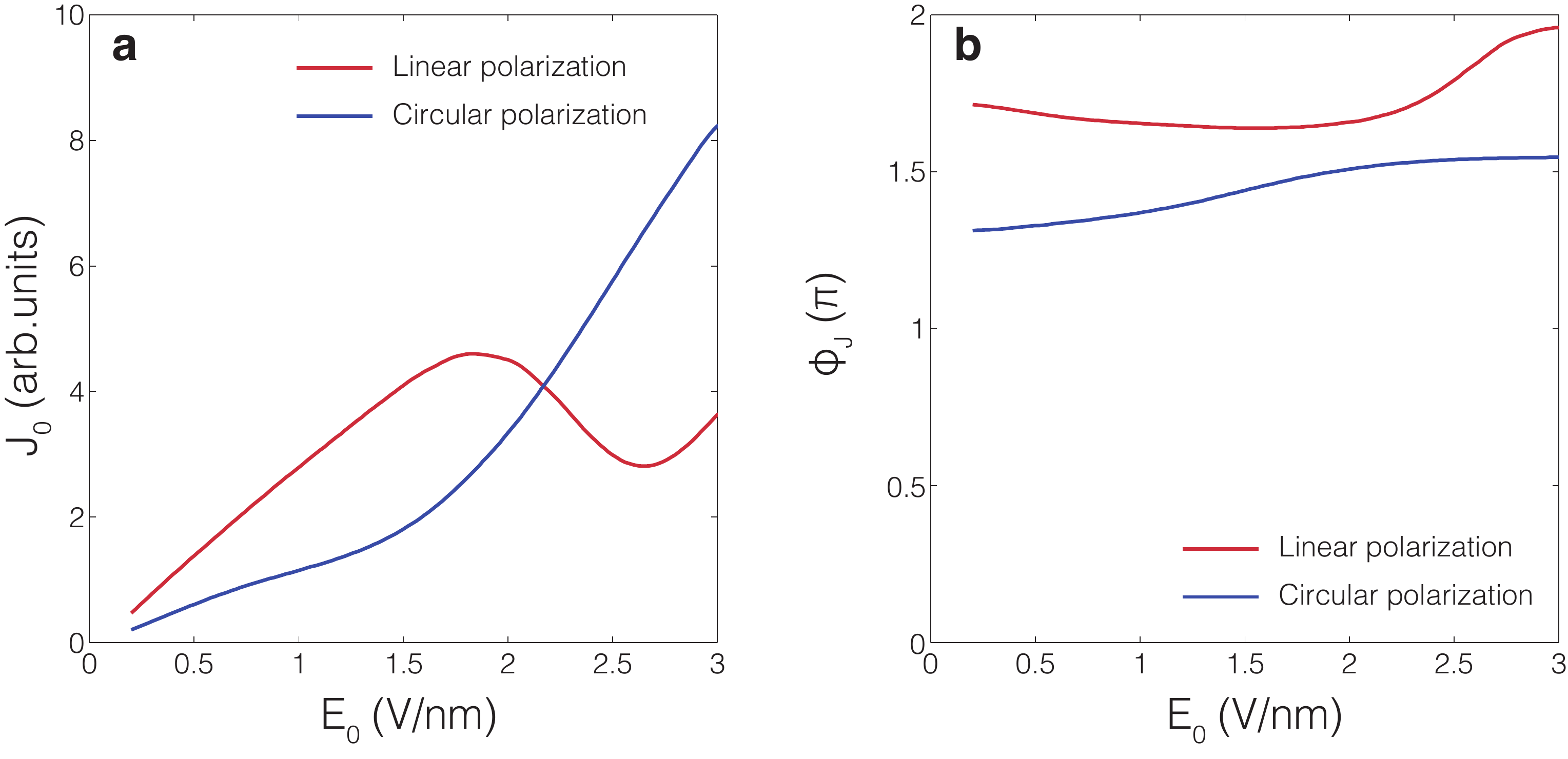}% Here is how to import EPS art
\caption{{\bf Numerically simulated results of charge transfer during the pulse} {\bf a.} $J_0$ and {\bf b.} $\phi_J$ as functions of the field amplitude $E_0$. }
\end{center}
\end{figure}

\begin{figure*}
\begin{center}
\includegraphics[width=14cm]{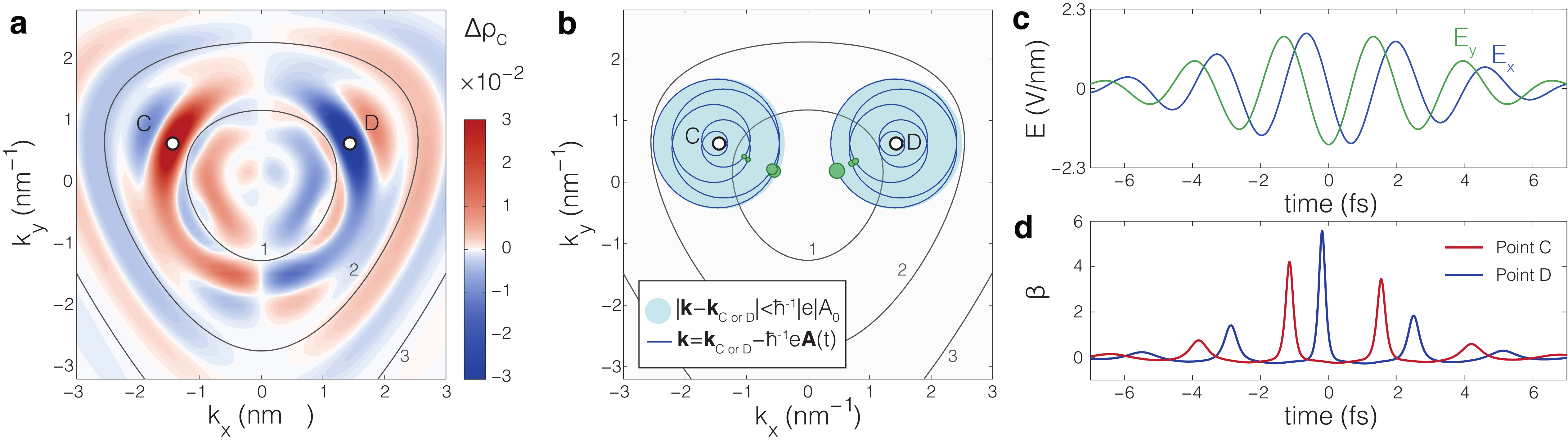}% Here is how to import EPS art
\caption{\label{Fig4} {\bf Relation between CEP-dependent current and electron trajectories under circularly polarized illumination.} {\bf a.} Difference $\Delta{\rho}_{\rm C}$ between conduction band populations after excitation by circularly polarized driving pulses with $\phi_{\rm CEP}=\pi/2$ and $-\pi/2$, shown as a function of the initial wave vector. $E_0$ equals to $2.3$ V/nm, and the peak field amplitudes of $x$ and $y$ component of the fields equal to $E_0/\sqrt{2}$.
Solid gray curves indicate wave vector values corresponding to resonant (multi-) photon absorption. The numbers of resonant photons are indicated by the numbers. {\bf b.} Electron trajectories under a circularly polarized pulse with $\phi_{\rm CEP}=\pi/2$, starting from two reciprocal points C and D. Green markers indicate the main transition events, which correspond to the peaks in $\beta(t)$. The size of the green markers is proportional to $|\beta(t)|$.
{\bf c} and {\bf d} Electric field waveforms and relative band coupling strength $\beta$ as function of time for the trajectories starting from C and D, respectively. Clearly, for circular polarization the asymmetry in band coupling can explain the CEP-dependence of the current. Intra-optical cycle quantum-path interference does not show up, which is why the direction of the CEP-dependent current does {\it not} change a function of $E_0$.}
\end{center}
\end{figure*}

\vspace{10 pt}
\noindent{\bf Estimation of the charge transfer during the interaction.}
From the temporal evolution of the conduction band population, one can also obtain the charge transfer due to the carrier motion during the pulsed excitation. Here the total current consists of the intraband and interband contributions. 
Here the time-dependent intraband current operator is
\begin{equation}
{\bf J}_{\lambda\lambda'}(t) = \delta_{\lambda\lambda'}\frac{e}{\hbar} \frac{\partial \varepsilon _{\lambda}({\bf k})}{\partial {\bf k}} \Big|_{{\bf k}={\bf k}_0 - \hbar^{-1} e {\bf A}(t)},
\end{equation}
where $\lambda$ and $\lambda'$ are the band indices ($c$ or $v$).
The intraband polarization operator ${\bf P}(t)$ is given in Eq.~\eqref{eq-DipoleMoment} (by substituting ${\bf k}={\bf k}_0 - \hbar^{-1} e {\bf A}(t)$).
The total time-dependent current ${\bf J}_{\rm tot}$ is
\begin{eqnarray}
{\bf J}_{\rm tot}(t) &=& \int_{{\rm BZ},\varepsilon ({\bf k}_0)>E_{\rm F}} d{\bf k}_0 \nonumber \\
&& \left[  \bra{\phi_{{\bf k}_0(t)}} {\bf J}(t) \ket{\phi_{{\bf k}_0(t)}}
+ \frac{\partial}{\partial t}\bra{\phi_{{\bf k}_0(t)}} {\bf P}(t) \ket{\phi_{{\bf k}_0(t)}} \right]. \nonumber \\
\end{eqnarray}
Note that the carrier multiplication does not occur on this short timescale. To compare with the experiments, we integrate this time-dependent current in time and the beam spot, and obtain $J_0$ and $\phi_J$ for this charge transfer. These results are plotted in Extended Data Figs.~3 a and b. Here we highlight that the $J_0$ and $\phi_J$ depend on $E_0$ and polarization quite differently from the experimentally observed behaviours (Figs.~2b and c). Therefore, this charge transfer during the pulse can be excluded for interpreting the experimental data.

\vspace{10 pt}
\noindent{\bf Origin of the CEP-dependent current for circularly polarized excitation.}
The experimentally observed laser-induced current also displays a CEP dependence for circularly polarized excitation (Fig.~2b). This CEP-dependent current comes from the CEP-dependent conduction band population after the laser pulse (Extended Data Fig.4~a), which can be interpreted on the basis of the asymmetric shape of the two-dimensional electron trajectories in reciprocal space (Extended Data Fig.~4b). The main difference between the $\phi_{\rm CEP} = \pi/2$- and the $\phi_{\rm CEP} = -\pi/2$-excitations is found at two initial reciprocal space points, C and D in Extended Data Figs.~4a and b. Extended Data Figures 4c and 4d show the electric field waveforms and $\beta(t)$ for electron trajectories initiated from C and D. Unlike in the case of linearly polarized excitation, only one positive peak and hence one transition event is found in $\beta(t)$ per optical cycle. Negative peaks are strongly suppressed. This is because the transition probability is maximized when the electron passes nearby the $K$ point as the dipole moment diverges at the $K$ point (Extended Data Figs.~2c and d). Therefore, the intra-optical-cycle interference that requires more than one transition per cycle cannot occur.

Unlike the case of linear polarization where CEP does not influence the magnitude of the $|\beta|$ peaks but rather spacings between them, the contrast of the magnitude of the $|\beta|$-peaks between different CEPs are prominent in the case of circular polarizations. This is the main mechanism for generating CEP-dependent current with circularly polarized laser pulses. The magnitude of $|\beta|$ behaves monotonically as a function of the field strength, and thus change of sign of the CEP-dependent current does not occur. Note that inter-cycle interference effects may happen, but the periodicity of the inter-cycle interference is always one optical cycle and is not affected by the CEP.

\end{document}